\documentclass[reprint,aps,prb,graphicx,superscriptaddress]{revtex4-2}
\usepackage{graphicx}
\usepackage{xcolor}
\usepackage[hidelinks]{hyperref}
\hypersetup{
    colorlinks,
    linkcolor={blue},
    citecolor={blue},
    urlcolor={blue}
}

\renewcommand\refname{References}

\draft 

\usepackage{mathtools}
\usepackage{physics}
\usepackage{enumitem}

\begin{document}
\renewcommand{\figurename}{Fig.}


\title{The role of nuclear spin diffusion in dynamic nuclear polarization of crystalline nanoscale silicon particles}

\author{Gevin von Witte}
 \affiliation{Institute for Biomedical Engineering, University and ETH Zurich, 8092 Zurich, Switzerland}
 \affiliation{Institute of Molecular Physical Science, ETH Zurich, 8093 Zurich, Switzerland}
\author{Konstantin Tamarov}
 \affiliation{Department of Technical Physics, University of Eastern Finland, 70210 Kuopio, Finland}
\author{Neva \c{S}ahin}
 \affiliation{Bilkent University, Ankara, Turkey}
\author{Aaron Himmler}
 \affiliation{Institute of Molecular Physical Science, ETH Zurich, 8093 Zurich, Switzerland}
\author{Vera Ganz}
 \affiliation{Institute of Molecular Physical Science, ETH Zurich, 8093 Zurich, Switzerland}
\author{Jani O. Moilanen}
 \affiliation{Department of Chemistry, Nanoscience Center, 
University of Jyväskylä, 40500 Jyväskylä, Finland}
\author{Vesa-Pekka Lehto}
 \affiliation{Department of Technical Physics, University of Eastern Finland, 70210 Kuopio, Finland}
\author{Grzegorz Kwiatkowski}
 \affiliation{Institute for Biomedical Engineering, University and ETH Zurich, 8092 Zurich, Switzerland}
\author{Sebastian Kozerke}
 \affiliation{Institute for Biomedical Engineering, University and ETH Zurich, 8092 Zurich, Switzerland}
\author{Matthias Ernst}
 \affiliation{Institute of Molecular Physical Science, ETH Zurich, 8093 Zurich, Switzerland}
 \email{maer@ethz.ch}

\date{\today}

\begin{abstract}
Hyperpolarized nanoparticles (NPs) offer high polarization levels with room temperature relaxation times exceeding half an hour.
In this work, we demonstrate that the achievable hyperpolarization enhancement and relaxation (decay) time at room temperature are largely independent of the particle size contrary to previous assumptions.
This is explained through first-principles spin-diffusion coefficient calculations and finite-element polarization simulations.
The simulated zero-quantum (flip-flop) line width governing the spin diffusion is found to agree with the experimentally accessible single-quantum (single spin flip, e.g. radio-frequency pulse) line width.
The transport of hyperpolarization from strongly hyperfine-coupled spins towards the bulk is most likelybelieved to be responsible for the slow polarization dynamics including long room temperature decay time. 
The line width and spin-diffusion simulations are extended to other cubic crystal structures and analytical expressions, which only require insertion of the gyromagnetic ratio, lattice constant, isotope abundance and measured spectral density distribution (nuclear line width), are fitted.
The presented simulations can be adjusted to study spin diffusion in other materials.
\end{abstract}

\maketitle


Recent advances in hyperpolarization have been stimulated by growing interest in its application to enhance the sensitivity of \textsuperscript{13}C organic compounds, specifically in the context of metabolic MR imaging \cite{jorgensen_hyperpolarized_2022}.
Small organic molecules, however, suffer from short spin-lattice relaxation times $T_1$ in solution on the order of few tens of seconds.
Short $T_1$ limits the time window available for imaging and imposes additional constraints on the sample transfer such as the need for fast dissolution \cite{ardenkjaer-larsen_increase_2003}, minimization of time for quality control and short shuttling distances between polarizer and imaging magnet.
Hyperpolarized (crystalline) nanoparticles can exhibit long $T_1$ times \cite{cassidy_vivo_2013,whiting_developing_2016,kwiatkowski_nanometer_2017,kwiatkowski_exploiting_2018,seo_efficient_2018,hu_hyperpolarization_2018,kim_29si_2021,von_witte_controlled_2024,kwiatkowski_direct_2018,waddington_phase-encoded_2019}, biocompatibility and sufficient \textit{in vivo} clearance \cite{zhu_biocompatibility_2012,ivanov_vivo_2012,vaijayanthimala_long-term_2012,croissant_degradability_2017,barone_pilot_2018,Santos2021,priyadarshni_nanodiamonds_2024,kwiatkowski_vivo_2024}.
Possible applications include background-free perfusion imaging not limited by the hyperpolarized $T_1$ time, targeted imaging (molecular targeting) and possible combination with therapeutics \cite{whiting_real-time_2015,whiting_developing_2016,Santos2021}.
    
Nanoparticles with a few tens of nanometers in diameter are preferred given their faster clearance to avoid \textit{in vivo} accumulation \cite{xu_size-dependent_2023}.
However, previous reports showed shorter $T_1$ times and lower enhancements for smaller particles \cite{aptekar_silicon_2009,whiting_developing_2016,rej_hyperpolarized_2015,kwiatkowski_direct_2018,boele_tailored_2020}. 
For diamond nanoparticles, the increased relaxation for smaller particles was associated with the larger ratio of surface to bulk defects \cite{rej_hyperpolarized_2015,boele_tailored_2020}.
Surface dangling bonds appeared rather inefficient for dynamic nuclear polarization (DNP) \cite{yoon_high-field_2019,kato_nanodiamond-based_2023} while causing strong relaxation. 
In contrast, bulk substitutional nitrogen defects (P1 or C-centers) can create DNP enhancements of several hundred \cite{rej_hyperpolarized_2015,kwiatkowski_direct_2018,boele_tailored_2020} even at room temperature \cite{shimon_large_2022,bussandri_p1_2023,nir-arad_nitrogen_2023}.
An increased relaxation by surface defects in nanodiamonds combined with an unchanged DNP injection from the bulk defects leads to a reduced steady-state polarization as shown using a rate-equation model \cite{von_witte_modelling_2023,von_witte_relaxation_2024} (cf. Sec.~S4 of the Supplementary Material \cite{US10nmSM} for a summary of the model).

For silicon, it was proposed that the long relaxation times ($>30$\,min) observed in microparticles resulted from their core-shell geometry \cite{dementyev_dynamic_2008,aptekar_silicon_2009,lee_decay_2011}.
According to their geometry, most of the polarization is stored in the slowly relaxing bulk, which is weakly connected via spin diffusion to the fast relaxing shell containing paramagnetic $P_b$ defects, which form naturally at the interface between the silicon core and the surface oxide.
The explanation for the long $T_1$ times of silicon came under scrutiny with measurements of room temperature $T_1\geq 40$\,min and enhancements $\varepsilon \geq 100$ in 50\,nm bulk particles \cite{kwiatkowski_nanometer_2017,kwiatkowski_exploiting_2018} as well as highly porous silicon (PSi, crystallite sizes between 10 and 60\,nm) particles \cite{von_witte_controlled_2024}.

In this work, we report on 20\,nm silicon particles with a room temperature $T_1\approx 40$\,min and DNP enhancements around 100 at 3.4\,T (3.4\,K).
This raises the question how to explain the long ambient relaxation times as the estimated spin-diffusion constant in silicon \cite{dementyev_dynamic_2008,hayashi_nuclear_2008} would suggest $T_1$ on the order of a few seconds.
We estimate the (isotropic) nuclear spin-diffusion coefficient $D$ from first principles for different cubic lattices.
The calculated $D$ is used to fit the build-up and room temperature decay with a finite-element approach \cite{pinon_measuring_2017,pinon_core-shell_2018} to show that the long experimental relaxation times are related to the slow nuclear spin diffusion near $P_b$ centers.

\section*{Nuclear spin diffusion}

Microscopically, nuclear spin diffusion results from the inter-nuclear dipolar couplings 
\begin{align}
    d_{ij} = \frac{\mu_0}{4\pi}\frac{\hbar \gamma_i \gamma_j}{r_{ij}^3}\frac{3\cos^2 \theta_{ij}-1}{2}, \label{eq:dipolarCoupling}
\end{align}
where $r_{ij}$ and $\theta_{ij}$ are the distance and angle between spins $i$ and $j$ with respect to the main magnetic field; $\gamma_i$ is the gyromagnetic ratio of spin $i$.
The dipolar coupling between the two spins leads to an exchange rate $W_{ij}$ which might be suppressed if the two spins have different resonance frequencies/energies (see below).
$W_{ij}$ can be computed by evaluation of the dipolar couplings on a lattice including a (normalized) spectral density function modulating the coupling \cite{lowe_density-matrix_1967,kubo_31p_1988,ernst_chapter_1998}.
In particular, we use \cite{ernst_chapter_1998}
\begin{align}
  W_{ij} = \frac{\pi}{2} d_{ij}^2 p_\mathrm{ZQ}(0),  \label{eq:Wij_ZQ}
\end{align}
where $p_\mathrm{ZQ}(0)$ is the normalized zero-quantum (ZQ) line evaluated at zero frequency.
This approach entails energy conserving flip-flops.
The ZQ line describes the energy level distribution of spin-spin flip-flops, which constitute the fundamental process of nuclear spin diffusion.
The factor $\frac{\pi}{2}$ arises from a Fermi's golden rule approach ($2\pi)$ combined with the notion that $-\frac{1}{2}d_{ij}(I_i^+ I_j^- + I_i^- I_j^+)$ (part B of the dipolar alphabet) is responsible for the flip-flops (providing an additional prefactor of 1/4).

In the following, we will discuss two different approaches how to calculate the spin diffusion coefficient
$D$. 

\textit{1. Nearest neighbor approach} \\
Assuming that spin diffusion is a process of nearest neighbors spin flip-flops, the spin diffusion coefficient $D$ can be calculated using \cite{bloembergen_interaction_1949,khutsishvili_spin_1966}
\begin{align}
    D_\mathrm{NN} = \frac{1}{2} W r_\mathrm{NN}^2 \label{eq:D_Bloembergen}
\end{align}
with $W$ being the spin flip-flop rate and $r_\mathrm{NN}$ the distance between nearest neighbor (NN) nuclear spins.
This approximation was used in previous estimates of the spin diffusion in silicon \cite{dementyev_dynamic_2008,hayashi_nuclear_2008} and relies on the experimentally measured single-quantum (SQ) line ($T_2^*$) describing the frequency/ energy distribution if a single spin is flipped (see below).
Specifically, the spin diffusion coefficient $D$ in the nearest neighbor approximation of cubic lattices is given by \cite{khutsishvili_spin_1966,hayashi_nuclear_2008}
\begin{subequations} \label{eq:nearestNeighbor}
    \begin{align}
        D_\mathrm{NN} &\approx \frac{\Delta\nu_{dd}}{30}r_\mathrm{NN}^2 \label{eq:nearestNeighbor_D}, \\
        r_\mathrm{NN} &\approx N_n^{-1/3} = \left(f\frac{8}{a^3}\right)^{-1/3}, \label{eq:nearestNeighbor_a}
    \end{align}
\end{subequations}
where $\Delta\nu_{dd}$ is the dipole-dipole contribution to the SQ line width \cite{hayashi_nuclear_2008}, $r_\mathrm{NN}$ is the statistical nearest neighbor distance  between \textsuperscript{29}Si atoms, $N_n$ is the spatial density of \textsuperscript{29}Si atoms, $f$ is the \textsuperscript{29}Si isotope abundance, and the 8 stems from the 8 lattice sites in the cubic conventional unit cell of silicon (diamond cubic structure) with the lattice constant $a=5.43$\,\AA.
The nearest neighbor distance $r_\mathrm{NN}$ may be modified by a factor of $\sim 0.55$ to account for the statistical occupation of the lattice (Poisson distribution, cf. appendix VII of \cite{chandrasekhar_stochastic_1943}) which would be relevant for not too high isotope abundances.
However, for high isotope abundances, this would result in an unrealistic nearest neighbor distance smaller than the distance between neighboring lattice sites.
We note that Eq.~\eqref{eq:nearestNeighbor_D} rather approximates a simple cubic mono-crystalline lattice as already discussed by Khutsishvili \cite{khutsishvili_spin_1966}.
A diamond cubic structure is a face-centered cubic (fcc) lattice with a diatomic basis.

\textit{2. Lattice approach} \\
A Taylor series expansion of the nuclear spin field yields the spin diffusion coefficient $D$ \cite{deng_nuclear_2005,sharma_enhancement_2019}
\begin{align}
    D_\mathrm{lat} = \sum_j \frac{1}{2} W_{ij}r_{ij}^2 \label{eq:D_sum}
\end{align}
with lattice site indices $i$ and $j$.

For a given central spin $i$, $W_{ij}$ is summed over all surrounding lattice sites \cite{deng_nuclear_2005,sharma_enhancement_2019}.
Combining Eqs.~\eqref{eq:Wij_ZQ} and~\eqref{eq:D_sum} gives 
\begin{align}
  D_\mathrm{lat} = \sum_j \frac{\pi}{4} d_{ij}^2 r_{ij}^2 p_\mathrm{ZQ}(0)  \label{eq:D_lattice} ~~~~~~. 
\end{align}

In this work, we focus on crystalline lattices consisting of a single atomic species (silicon).
Hence, we can approximately calculate the ZQ line using a three spin model: (i) a central spin, (ii) a second spin which we designate as the target spin for the spin flip-flop and (iii) a third background spin modifying the energy levels of the first two spins.
The second and third spin (target and background spins) take random positions with respect to the central spin and, in principle, we need to sum over all possible three spin systems.

In this three spin system, we study the zero and single-quantum (ZQ and SQ) processes of the system.
The SQ line describes the experimentally measured line width with a single spin flipped by a RF pulse.
The ZQ line describes the spin exchange processes with a net-zero change of spin angular momentum (spin flip-flop) occurring during nuclear spin diffusion. 
We note that the ZQ and SQ lines can have different values. 
Therefore, calculating the spin diffusion coefficient $D$ based on the experimentally observed SQ line ($\sim 1/T_{2,\mathrm{SQ}}$) might give erroneous results for the spin diffusion coefficient. 

In this work, we compute the SQ and ZQ line with an approach based on Van Vleck moments \cite{van_vleck_dipolar_1948,abragam_principles_1961,mehring_principles_1983}. 
We assume a Gaussian line such that the SQ and ZQ line widths for the full width at half maximum (FWHM) are given by $\mathrm{FWHM} = 2\sqrt{2 \mathrm{ln}(2) M_2}$ with $M_2$ being the second moment.
The lengthy expressions for $M_2$ of the three spin model are provided in a Mathematica (Wolfram Research, USA) notebook together with the experimental data (cf. Materials \& Correspondence section).
However, it should be noted that the expressions for $M_2$ of SQ and ZQ line only contain $d_{ij}^2$, which will be used to define an effective cut-off distance to accelerate the lattice calculations.
Furthermore, the relevant part of the dipolar coupling for $M_2$ is $2d_{ij}I_i^z I_j^z$ (part A of the dipolar alphabet), giving an additional factor of 2 in the ZQ/SQ line calculation.

\section*{Methods}

\subsection*{Samples and characterization}
The samples were purchased from US Nano Research (USA) and used without further modification.
For the 50\,nm sample, the previously reported characterization of the sample was used \cite{kwiatkowski_nanometer_2017,kwiatkowski_exploiting_2018} .
For the 20\,nm particles, the average particle size (APS) was measured with a transmission electron microscope (TEM) \cite{kwiatkowski_exploiting_2018}.
The 20\,nm particles were characterized with X-band electron paramagnetic resonance (EPR,  Magnettech MiniScope MS5000,
Bruker Corp.) following the experimental approach and analysis presented in \cite{von_witte_controlled_2024}.
In particular, the density of paramagnetic centers was calculated with a reference TEMPO sample and the EPR spectrum was fitted with a combination of anisotropic $P_b^\mathrm{111}$ and isotropic $P_b^\mathrm{iso}$ centers using EasySpin 6.0.6. 
More details about the EPR measurements and analysis can be found in \cite{von_witte_controlled_2024}.

The Si crystalline sizes were measured with X-ray powder diffraction (XRPD, D8 Discover, Bruker Corp.) in Bragg–Brentano geometry.
Si crysalline sizes were then calculated from the diffraction pattern using Rietvield refinement method in TOPAS\textsuperscript{\textregistered} 4.6 software taking into account the instrumental broadening.

\subsection*{Dynamic nuclear polarization (DNP)}

The DNP profiles and build-up curves were measured using home-built polarizers operated at 3.4 and 7\,T \cite{jahnig_dissolution_2017} and a temperature of 3.4\;K unless otherwise stated. 
The 3.4\,T system was equipped with an OpenCore NMR \cite{takeda_highly_2007,takeda_opencore_2008,takeda_chapter_2011} (142\;MHz \textsuperscript{1}H Larmor frequency) spectrometer and the 7\,T with a Bruker Avance III console (Bruker BioSpin, Switzerland, 299\;MHz \textsuperscript{1}H Larmor frequency).
Both set-ups were equipped with a 200\,mW MW source (from ELVA-1, Estonia for 3.4\,T and Virginia Diodes (VDI), USA for 7\,T) unless otherwise stated.
DNP profiles at 3.4\,T (cf. Sec.~S2 of the Supplementary Material \cite{US10nmSM}) were performed with a 400\,mW VDI source.
The 7\,T set-up was equipped with an in-house electroplated low-loss wave guide yielding approximately doubled MW power at the sample space compared to a stainless steel waveguide (around 65\;mW at the sample) \cite{himmler_electroplated_2022}. 
The MW power was set to the maximum output unless explicitly stated. 
The MW was frequency-modulated to further improve the polarization (cf. Sec.~S2 of the Supplementary Material \cite{US10nmSM}) \cite{kwiatkowski_nanometer_2017,kwiatkowski_exploiting_2018}: a sawtooth modulation with 1\,kHz (10\,kHz) frequency and $\sim$150 (300)\,MHz bandwidth at 3.4 (7)\,T was used.
The other details of the set-up are described elsewhere \cite{jahnig_dissolution_2017,himmler_electroplated_2022}. 
The experiments at 3.4\,T were performed with small flip angles ($\sim$1.5\textdegree) and long repetition times (20\,min) while at 7\,T about 7\textdegree\; pulses every six minutes were used.
Perturbations of the signal intensity by the monitoring radio-frequency (RF) pulses were corrected \cite{von_witte_modelling_2023}.

The results reported in this work were measured together with porous silicon (PSi) experiments reported in \cite{von_witte_controlled_2024}.
In particular, both reports rely on the same thermal-equilibrium measurements (see below) and the identical set-ups allowed for a direct comparison of the absolute polarizations/enhancements. 
Measuring the thermal equilibrium for silicon is complicated by the long relaxation times and the low signal.
Therefore, we used the averaged signal measured with a fully \textsuperscript{29}Si labeled sample at room temperature, following previous work \cite{kwiatkowski_nanometer_2017,kwiatkowski_exploiting_2018}.
At 3.4\,T, the averaged signal of a saturation recovery experiment with a single data point (after $\sim$3.8\,h) was extrapolated to infinite time with a $T_1$ from a hyperpolarized decay experiment.
At 7\,T, multiple thermal build-up (saturation recovery) experiments were averaged and the fitted value taken for the thermal signal.

The room temperature relaxation time was measured with a 9.4\,T small animal magnetic resonance imaging (MRI) scanner (Bruker BioSpin, Germany) employing a solenoid coil wound around the sample cup \cite{kwiatkowski_nanometer_2017,kwiatkowski_exploiting_2018}.
The polarization before the room temperature relaxation relied on a different cryostat inset \cite{batel_multi-sample_2012} together with a 1\,W MW amplifier (QuinStar, USA) inserted after the frequency-modulated MW source.
Due to the absence of a flip angle calibration, it was not possible to correct the room temperature relaxation measurement for the perturbations by the RF pulses.

All data processing was performed with in-house developed MATLAB (MathWorks Inc., USA) scripts. 
Uncertainties in the processed experimental data are expressed as 95\,\% fit intervals.
Experimental instabilities such as slight changes in the MW output or temperature fluctuations as well as uncertainties in the thermal equilibrium measurement were considered negligible. 

\subsection*{ZQ line and spin-diffusion simulations}

To accelerate the ZQ and spin-diffusion simulations, a spatial cut-off distance for the respective couplings ($d_{ij}^2$ and $d_{ij}^2r_{ij}^2$) was set on 100 randomly generated lattices per isotope abundance.
The simulation box for the ZQ line and spin diffusion spanned 30 lattice constants (around 150\,\AA) in every direction from the central unit cell giving a cube of $61^3$ unit cells with 8 sites per unit cell.

In the ZQ line simulations, the identified cut-off distances (cf. Fig.~\ref{fig:Fig3} for silicon and Fig.~S10 of the Supplementary Material \cite{US10nmSM} for the generalized spin-diffusion simulations of cubic lattices) were used to accelerate the calculations with a moderate sacrifice in accuracy as at least 95\% of the relevant coupling was chosen to be within the cut-off.
The exact cut-off values are given in the simulation files and are available online (cf. Materials \& Correspondence section).
The secondary spins (target spins for flip-flop with the central spin) consist of all spins (except the central spin) for the ZQ and SQ simulations within one cut-off distance.
For the background spins, all spins (except the chosen secondary and central spins) within twice the cut-off distance contribute in the simulations. 
Averaging over all secondary spins was applied in the SQ and ZQ line calculations.
For the ZQ/ SQ and spin-diffusion simulations, around 1600 orientations and 100 randomly generated lattices were averaged for each isotope abundance.
The dipolar cut-off, ZQ line and spin-diffusion simulations were carried out with in-house developed Matlab scripts.

\subsection*{Finite element build-up and decay simulations}

\begin{figure}[ht]
	\centering
	\includegraphics[width=0.7\linewidth]{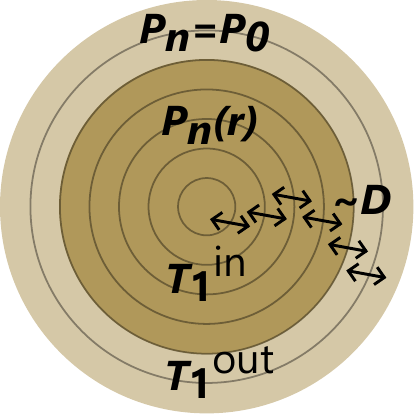}
	\caption{Sketch of the finite element build-up simulations with the outer (shell) and inner (core) particle parts characterized by different relaxation times and polarization levels.
    In the inner part of the particle, the polarization depends on the radial segment.
    During the build-up, the polarization in the outer part is fixed at the steady-state polarization.
    The nuclear spin diffusion spreading the hyperpolarization is assumed to be isotropic.}
	\label{fig:Fig1}
\end{figure}

The finite element build-up and decay simulations were adopted from published work \cite{pinon_measuring_2017,pinon_core-shell_2018} and implemented in Matlab.
Fig.~\ref{fig:Fig1} sketches the approach.
A particle radius of 10\,nm (25\,nm) was assumed with an outer (shell) part of 3\,nm thickness containing the defects responsible for DNP.
The thickness of the outer shell is motivated by the difference between the APS of the 50\,nm particles and crystalline size (XRD) of 43\,nm.
The hyperpolarization build-up was modeled by setting the nuclear polarization in the outer part (shell) to be equal to the experimentally measured steady-state value, since this value is constantly replenished by DNP from the electrons.
The particle was discretized in 1000 radial elements.
The spin diffusion value was chosen based on the ZQ line and spin-diffusion simulations. 
For simplicity, the spin diffusion in the outer part of the particles and in its core was assumed identical.
Possible quenching of spins and a different spin concentration in the outer shell were ignored in the absence of a good estimate.
The nuclear polarization relaxation times $T_\mathrm{1,in}$ and $T_\mathrm{1,out}$ of the inner (core) and outer (shell) part of the particle were fitted in a least-squares grid search while all other parameters were fixed.
We assume a homogeneous relaxation time $T_\mathrm{1,out}$ for the outer shell that absorbs a number of effects that are inhomogeneous over the outer shell: (i) the dependence of the paramagnetic relaxation as a function of the distance (and orientation) to the paramagnetic center; (ii) the dependence of the spin-diffusion rate constant on the distance (and orientation) to the paramagnetic center due to frequency detuning.

\section*{Results and Discussion}

\begin{figure*}[ht]
	\centering
	\includegraphics[width=\linewidth]{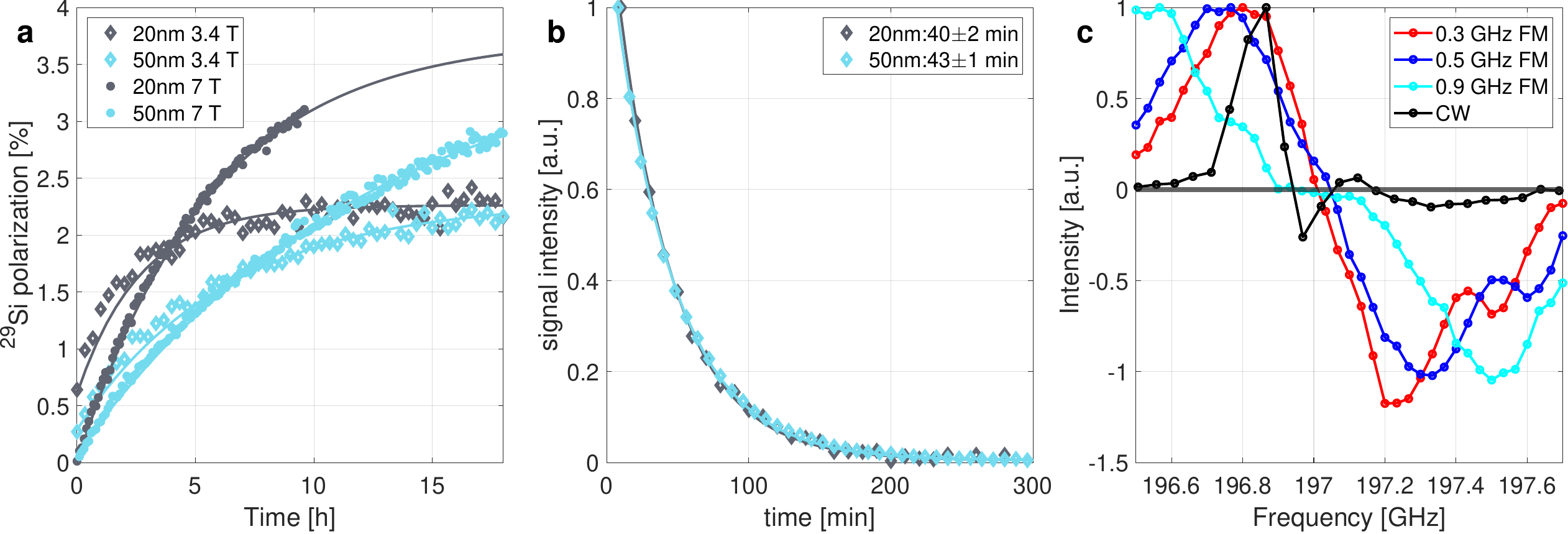}
	\caption{\textbf{(a)} Build-up comparison of 20 and 50\,nm particles at 3.4 and 7\,T with MW modulation. 
    \textbf{(b)} Decay at room temperature in a 9.4\,T scanner after polarization for around 24\,h at 3.4\,T. 
    \textbf{(c)} DNP profiles of the 20\,nm particles with different frequency modulation (FM) widths (cf. Sec.~S2  of the Supplementary Material \cite{US10nmSM} for more details).
    For the continuous wave (CW) MW irradiation, only a very weak negative DNP enhancement can be observed. 
    With frequency modulation, the positive and negative parts of the DNP profile have similar amplitude and shape.
    A comparison with the DNP profile of the 50\,nm particles, an EPR profile and further discussion is given in Sec.~S2 of the Supplementary Material \cite{US10nmSM}.
    The 20 and 50\,nm samples show identical enhancements and room temperature relaxation while differing in their build-up times and DNP profiles.}
	\label{fig:Fig2}
\end{figure*} 

\subsection*{Experiment}

Figure~S1 of the Supplementary Material \cite{US10nmSM}, shows the transmission electron microscope (TEM) image of the 20\,nm particles and the particle size distribution, giving an average particle size (APS) of 20$\pm$12\,nm.
The estimated crystalline sizes from XRPD were $25.8\pm 0.4$\,nm with a indication for an oxide phase (Fig.~S2 of the Supplementary Material \cite{US10nmSM}).

The comparison of the hyperpolarization build-ups corrected for the perturbations of the monitoring RF pulses \cite{von_witte_modelling_2023} is displayed in Fig.~\ref{fig:Fig2}a for the 20\,nm and 50\,nm particles ($43$\,nm mean crystalline size \cite{kwiatkowski_exploiting_2018}) at 3.4 and 7\,T (3.4\,K at both fields).
Both samples reached the same fitted nuclear steady-state polarizations (enhancements $\varepsilon$, cf. Fig.~S7 of the Supplementary Material \cite{US10nmSM}) at the two fields: 2.3\% ($\varepsilon = 112$) and 3.7\% ($\varepsilon = 89$) at 3.4 and 7\,T, respectively.
The polarizations/enhancements are similar to those reported for porous silicon (PSi) particles \cite{von_witte_controlled_2024}, although the best performing PSi samples provided higher enhancements at 3.4\,T and lower at 7\,T (both data sets are calibrated with the same thermal equilibrium measurements, see Methods).
We note the good reproducibility of DNP results and thermal equilibrium measurements as the 50\,nm sample was reported with an enhancement of 97 at 3.4\,T in independent experiments \cite{kwiatkowski_exploiting_2018}.
At both fields, the 20\,nm particles show a faster build-up time (2.6\,h and 5.6\,h at 3.4\,T and 7\,T, respectively) than the 50\,nm sample (6.3\,h and 12.4\,h). 
This difference is discussed in Sec.~S4 of the Supplementary Material \cite{US10nmSM} in terms of rate equation parameters \cite{von_witte_modelling_2023}, which show a (partial) limitation of the achievable polarization levels by the MW-induced relaxation enhancement \cite{von_witte_relaxation_2024,von_witte_controlled_2024}, i.e., the MW irradiation reduces the electron polarization at the irradiation frequency, which increases the rate of triple spin flips (electron-electron-nuclear flip-flop-flips) causing nuclear relaxation.

Fig.~\ref{fig:Fig2}b compares the room temperature relaxation of the 20 and 50\,nm samples recorded at a 9.4\,T small animal MRI.
This data is not corrected for the perturbations of the monitoring RF pulses \cite{von_witte_modelling_2023} as the exact flip angles of the different experiments are not known (cf. Methods).
The 20 and 50\,nm particles show nearly identical relaxation times of 40 and 43\,min at room temperature.

Differences between the two samples are evident in the DNP profiles, which will be summarized below and discussed in more detail in Sec.~S2 of the Supplementary Material \cite{US10nmSM}.
The 50\,nm sample shows a nearly symmetric DNP profile \cite{kwiatkowski_exploiting_2018} (cf. Fig.~S5 of the Supplementary Material \cite{US10nmSM}) even without MW frequency modulation, in agreement with other samples \cite{dementyev_dynamic_2008,kwiatkowski_exploiting_2018,von_witte_controlled_2024}.
In contrast, the DNP profile of the 20\,nm sample is asymmetric (cf. Fig.~\ref{fig:Fig2}c) with the DNP lobe of negative enhancement nearly vanishing for the continuous wave (CW) MW irradiation.
Furthermore, instead of a wide negative DNP lobe, a weak narrow positive DNP enhancement is observed around 197.2\,GHz in Fig.~\ref{fig:Fig2}c.
A similar asymmetry although possibly for different reasons has been observed in \cite{atkins_synthesis_2013} for nanoparticles with different surface terminations.
If MW irradiation is frequency modulated, the DNP profile of the 20\,nm particles appears nearly symmetric.
The asymmetry of the continuous wave (CW) DNP profile and its nearly symmetric shape with frequency modulation (FM) are not understood as of now and require further study.

In summary, the 20\,nm particles show identical enhancements and room temperature $T_1$ compared to 50\,nm particles and \textmu m particles \cite{kwiatkowski_exploiting_2018,kwiatkowski_vivo_2024}.

\subsection*{Simulation of silicon}


\subsubsection*{Cut-off distance of nuclear couplings}

\begin{figure}[ht]
	\centering
	\includegraphics[width=\linewidth]{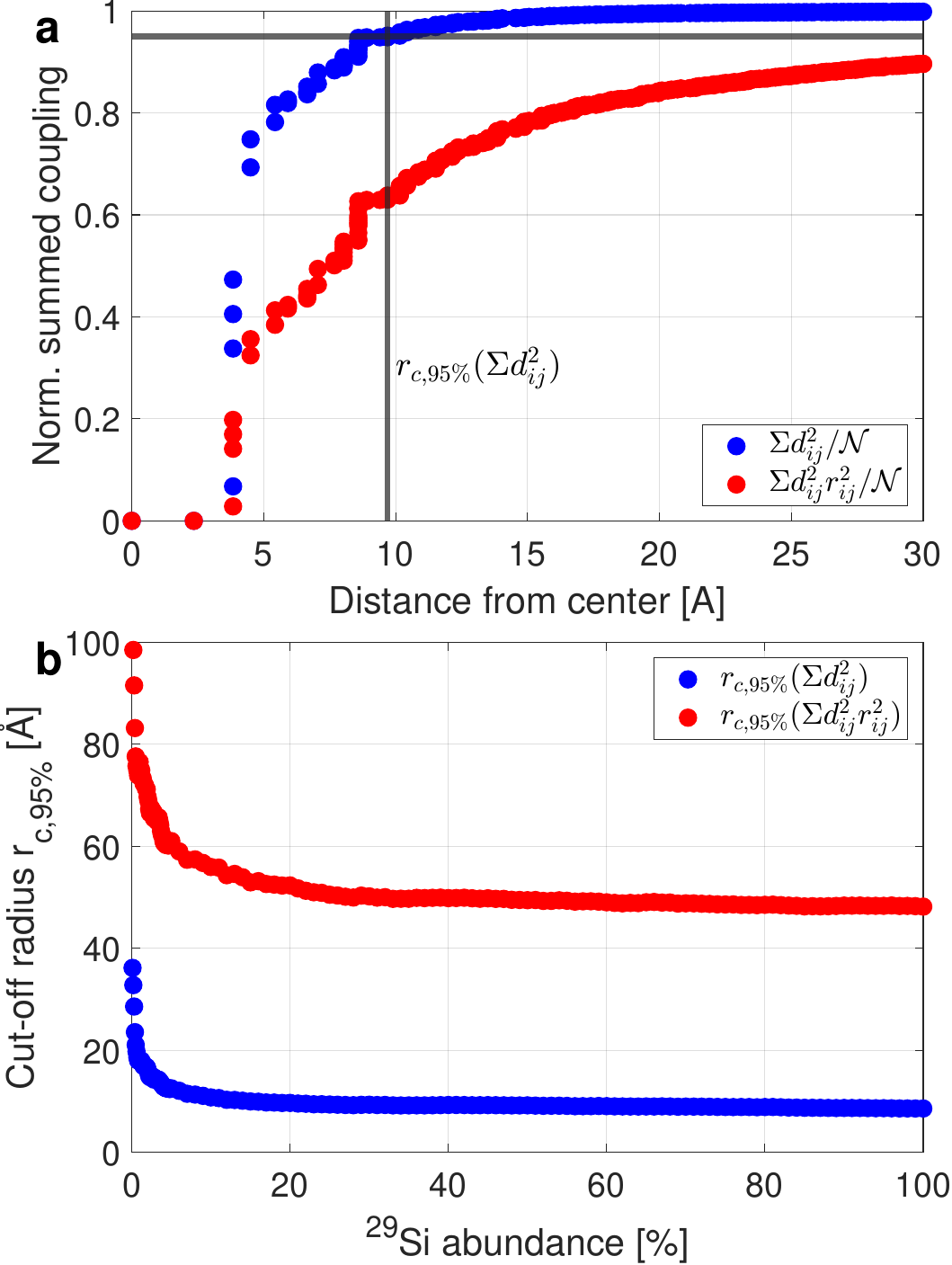}
	\caption{\textbf{(a)} Normalized inter-nuclear distance dependence of the sum over $d_{ij}^2$ and $d_{ij}^2 r_{ij}^2$ calculated on a 30\% \textsuperscript{29}Si lattice. 
    The former describes the nuclear dipolar flip-flops, which governs the SQ and ZQ line widths while the latter describes spin diffusion (cf. Eq.~\eqref{eq:D_lattice}).
    The slow convergence of $\sum_j d_{ij}^2 r_{ij}^2$ indicates the importance of distant spins, i,e. a few nanometers away, for spin diffusion.
    The sums are normalized over all spins within the simulation box of around 150\,\AA. 
    \textbf{(b)} Cut-off distance $r_\mathrm{c, 95\%}$ to contain 95\% of total $d_{ij}^2$ and $d_{ij}^2 r_{ij}^2$ by \textsuperscript{29}Si abundance (cf. panel (a)).
    These cut-off distances will be used to accelerate the SQ/ ZQ line and spin-diffusion simulations as only a limited number of spins needs to be considered to obtain accurate results.}
	\label{fig:Fig3}
\end{figure}

Fig.~\ref{fig:Fig3}a describes the inter-nuclear distance dependence of the sum over $d_{ij}^2$ and $d_{ij}^2 r_{ij}^2$ normalized to the summed interaction over all spins within $\sim$150\,\AA\, of a chosen central spin.
The dipolar couplings were calculated on a 30\% \textsuperscript{29}Si lattice with the magnetic field implicitly assumed along the (100) direction of silicon's diamond cubic structure.
Please note that we use the term ''lattice site'' for the crystallographic lattice throughout this work and reserve ''nearest neighbor'' for the nearest \textsuperscript{29}Si to a given \textsuperscript{29}Si atom.
A few aspects of Fig.~\ref{fig:Fig3}a should be noted: The coupling to the nearest neighbor is zero as this neighbor is at the magic angle. 
The groups of next-nearest and next-next-nearest neighbors can have a similar coupling to the central spin depending on their position in the randomly generated lattice.
For \textsuperscript{29}Si spins around 10\,\AA\, (or about two lattice constants) away from the central spin, the sums over $d_{ij}^2$ and $d_{ij}^2 r_{ij}^2$ become nearly continuous and converge to a finite value.
Around 95\% of the total coupling of $d_{ij}^2$ to the central spin originates from spins less than two lattice constants away.
$d_{ij}^2 r_{ij}^2$ shows a slower spatial convergence and only around 60\% of the total coupling is contained within two lattice constants.
We remark that spin flip-flops at several lattice constants away are essential to accurately describe spin diffusion with its $d_{ij}^2 r_{ij}^2$ scaling (cf. Eq.~\eqref{eq:D_lattice}).

Figure~\ref{fig:Fig3}b shows the cut-off distances for 95\,\% of the total coupling contained within the cut-off for different \textsuperscript{29}Si isotope abundances.
For low abundances, a larger cut-off value is necessary than for dense \textsuperscript{29}Si spins.
Thus, for low abundance lattices, e.g. 4.7\,\% natural \textsuperscript{29}Si abundance, flip-flops between spins more than 50\,\AA\, away from each other still contribute to the total spin diffusion.
We note that such a large distance for nuclear flip-flops with respect to the 20\,nm average particle size might raise problems with the use of a coarse-grained macroscopic spin diffusion coefficient $D$.
However, we will estimate $D$ and use it to fit the build-ups and decays as the simulations seem to be independent of the exact value of $D$ as will be discussed below.

\subsubsection*{Nuclear SQ and ZQ line widths}

\begin{figure}[ht]
	\centering
	\includegraphics[width=\linewidth]{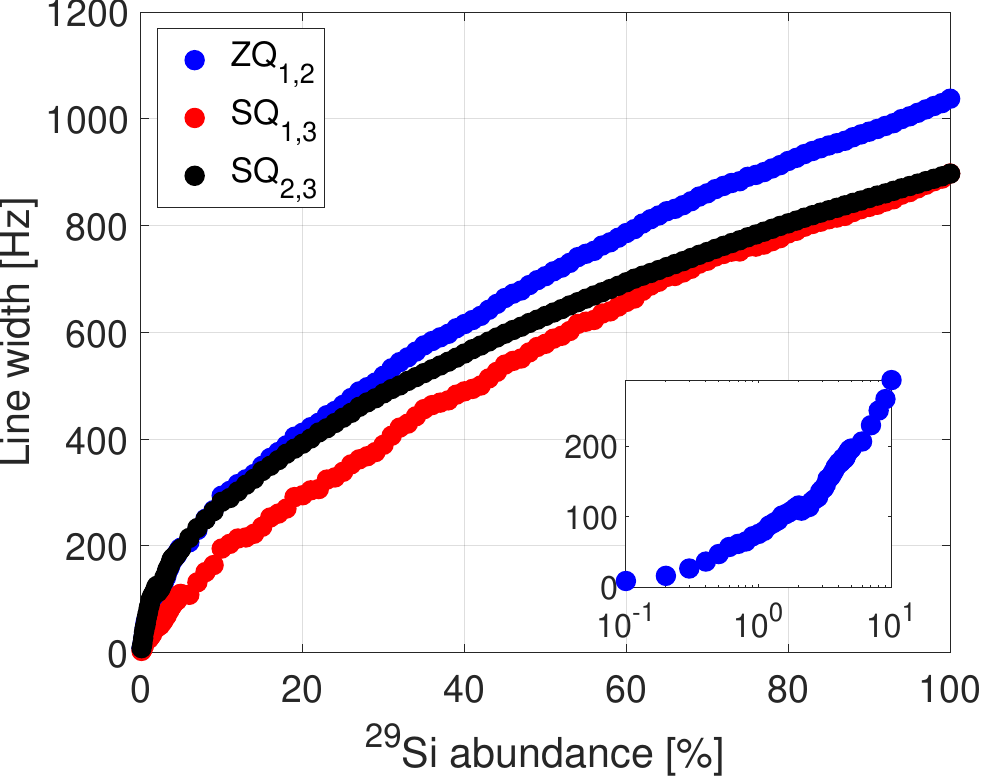}
	\caption{Simulated SQ and ZQ line widths for the three spin system. 
    The inset shows the dependence of the ZQ line width for low \textsuperscript{29}Si abundances. 
    The ZQ and SQ have similar line widths, justifying the use of experimental SQ lines to estimate the spin diffusion \cite{hayashi_nuclear_2008,dementyev_dynamic_2008}.
    Compared to experimental dipolar SQ line widths, e.g. around 20, 120 and 1700\,Hz for 1, 10 and 100\% \textsuperscript{29}Si abundance, the simulated values are too large for low \textsuperscript{29}Si abundances and too small for high. 
    This is attributed to the assumed Gaussian line shape to convert the $M_2$ into a line width (cf. Nuclear spin diffusion section) in the simulation compared to the more complicated line shapes observed experimentally \cite{hayashi_nuclear_2008}.
    The small bumps at 2\% \textsuperscript{29}Si abundance are ascribed to changes in the cut-off distance (cf. Methods).}
	\label{fig:Fig4}
\end{figure}

The simulation results of the SQ and ZQ line widths of the three spin model evaluated on randomly generated diamond cubic structures for varying \textsuperscript{29}Si abundances are shown in Fig.~\ref{fig:Fig4}.  
The SQ and ZQ line widths are a result of powder averaging over nearly 1600 orientations with considerable differences depending on the angle as shown in Fig.~S9 of the Supplementary Material \cite{US10nmSM}.
Crucially, the ZQ and SQ lines have very similar widths, which means that experimentally obtained values for the dipolar SQ line width can be used to estimate the spin diffusion in silicon. 

Compared to the experimental dipolar SQ line widths reported in \cite{hayashi_nuclear_2008} e.g. around 20, 120 and 1700\,Hz for 1, 10 and 100\% \textsuperscript{29}Si abundance, the simulated line widths are too large for low, accurate for intermediate and too small for high \textsuperscript{29}Si abundances.
For low abundances, the experimental NMR line is close to a Lorentzian while for intermediate \textsuperscript{29}Si concentrations a nearly single Gaussian line is observed \cite{hayashi_nuclear_2008}. 
For a highly \textsuperscript{29}Si labeled sample, the line can be approximated by multiple Gaussians \cite{hayashi_nuclear_2008}.
The latter result is consistent with a fully \textsuperscript{29}Si labeled powder sample measured in our lab under identical settings (data not shown).
The labeled sample is mostly used to measure the \textsuperscript{29}Si thermal equilibrium signal (see Methods).
In the simulations, we always assumed a Gaussian line, which leads to an overestimation of the SQ/ZQ line width for low and an underestimation for high \textsuperscript{29}Si abundances.

\subsubsection*{Spin diffusion coefficient}

\begin{figure}[ht]
	\centering
	\includegraphics[width=\linewidth]{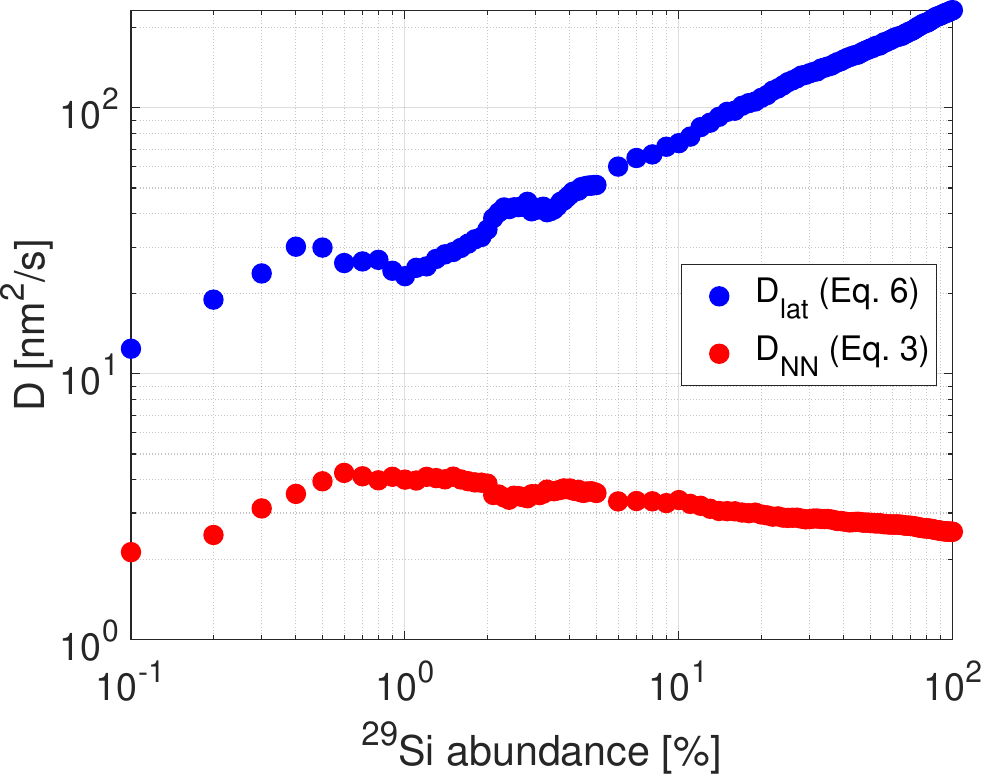}
	\caption{Spin diffusion coefficient $D$ extracted from the nearest neighbor approximation as given in Eqs.~\eqref{eq:nearestNeighbor} showing nearly identical values to \cite{hayashi_nuclear_2008}. 
    Extracting the spin diffusion coefficient with a lattice calculation of Eq.~\eqref{eq:D_lattice} gives much larger spin diffusion values. 
    The difference between the two approaches might result from the contributions of more distant nuclear spins (cf. main text and Fig.~\ref{fig:Fig3}).
    The small bumps at 2\,\% \textsuperscript{29}Si abundance are from the change in cut-off distance (cf. Methods and Fig.~\ref{fig:Fig4}).}
	\label{fig:Fig5}
\end{figure}

Converting the simulated ZQ line width into a spin-diffusion coefficient $D$ with the nearest neighbor approximation as given in Eqs.~\eqref{eq:nearestNeighbor} results in similar values for $D_\mathrm{NN}$ (cf. Fig.~\ref{fig:Fig5} compared to \cite{hayashi_nuclear_2008}, e.g. for 4.7\% natural abundance 1.7\,nm\textsuperscript{2}/s \cite{hayashi_nuclear_2008} and 3.6\,nm\textsuperscript{2}/s in our simulations. 
The difference between the two values might be explained by the assumed Gaussian line shape in the simulations, leading to too large line widths for low \textsuperscript{29}Si abundances (see above).
Qualitatively, the dependence of $D_\mathrm{NN}$ in the nearest neighbor model (Eqs.~\eqref{eq:nearestNeighbor}) with \textsuperscript{29}Si abundance shows a maximum for less than one percent and a moderate decrease for high abundances.
This is in contrast to experiments \cite{hayashi_nuclear_2008} and previous simulations on a different lattice \cite{nolden_simulation_1996} with both predicting a moderate monotonic increase.
The discrepancy might result from an overestimation of the line width for low and an underestimation for high abundances (see above).

Estimating the nuclear spin diffusion based on a lattice approach given by Eq.~\eqref{eq:D_lattice} yields $D_\mathrm{lat} \approx 51$\,nm\textsuperscript{2}/s for 4.7\% natural abundance.
Considering the rather slow convergence of the summed (cumulative) coupling to the central spin (cf. Fig.~\ref{fig:Fig3}), the estimated spin diffusion is an order of magnitude larger than for the nearest neighbor model (cf. Eqs.~\eqref{eq:nearestNeighbor}) as shown in Fig.~\ref{fig:Fig5}.

To support the results of the lattice approach, which give an order of magnitude larger natural abundance spin diffusion coefficient compared to the more commonly used nearest neighbor approach, we discuss the relaxation experiments by Lee et al. \cite{lee_decay_2011}.
The relaxation experiments in \cite{lee_decay_2011} assumed smaller particles in the simulations than experimentally used to match experiment and theory, i.e. the particles had a diameter of 5\,\textmu m while in the simulations a diameter of 700\,nm was used.
If spin diffusion would be faster than assumed in the relaxation simulations, this could explain the need for reduced particle sizes in the simulations, possibly supporting the validity of the larger spin diffusion coefficient of the lattice approach (Eq.~\eqref{eq:D_lattice}).
This might suggest that spin diffusion is faster than estimated in the nearest neighbor model for silicon \cite{hayashi_nuclear_2008,dementyev_dynamic_2008}.
Build-up and decay finite element simulations shown below cannot discriminate between the two spin diffusion coefficients as both lead to reasonable results. 
Therefore, further studies are required to understand how to accurately simulate the spin diffusion coefficient from first principles.

\subsubsection*{Finite-element nanoparticle simulations}

\begin{figure*}[ht]
	\centering
	\includegraphics[width=\linewidth]{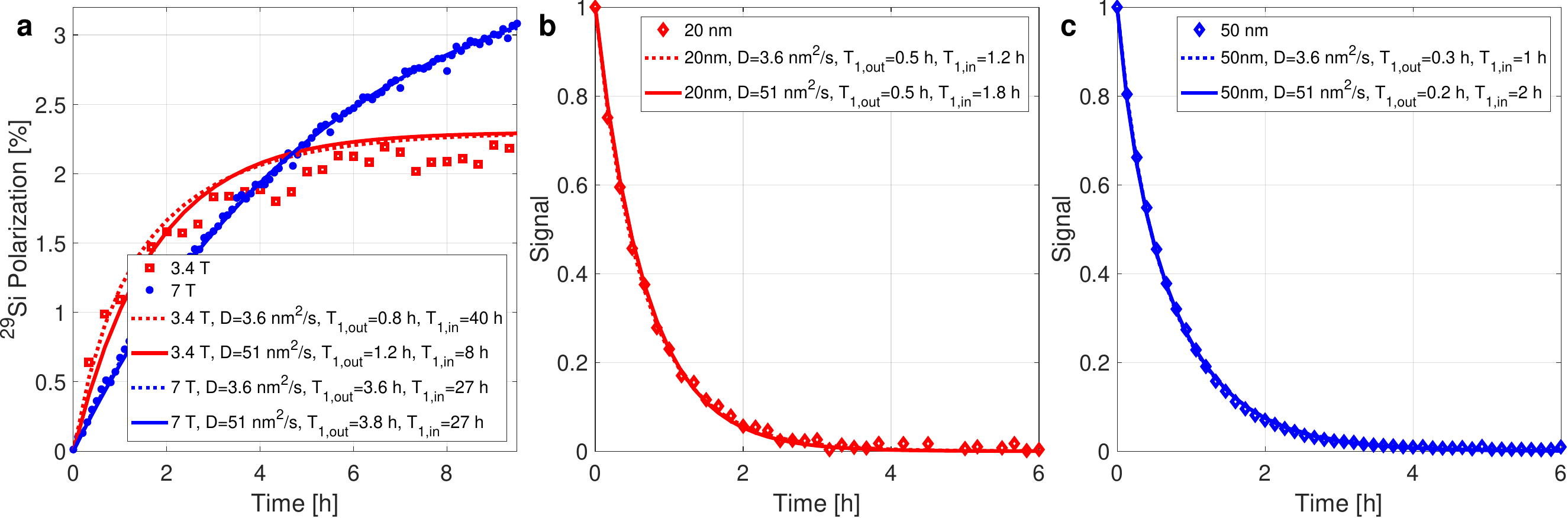}
	\caption{\textbf{(a)} Experimental and simulated build-ups at 3.4 and 7\,T (cf. Fig~\ref{fig:Fig2}a) of the 20\,nm particles.
    \textbf{(b, c)} Experimental and simulated room temperature decays of the 20 and 50\,nm particles at 9.4\,T (cf. Fig~\ref{fig:Fig2}b).
    The simulated build-ups and decays are characterized by a slow relaxation in the inner (core, $T_\mathrm{1,in}$) part of the particle and faster relaxation in the outer (shell, $T_\mathrm{1,out}$) part of the particle. 
    The choice of the spin diffusion coefficient $D$ has little effect on the fits.
    The different $T_1$ values at 3.4\,T are attributed to few, low SNR data points, resulting in an insensitive least squares residue (cf. Fig.~S12 of the Supplementary Material \cite{US10nmSM}).}
	\label{fig:Fig6}
\end{figure*}

Spin diffusion times of spherical 20\,nm particles ($D \approx 3.6$ or 51\,nm\textsuperscript{2}/s) are shorter than one minute, which is much shorter than the hours long build-up times. 
To better understand this slow hyperpolarization build-up, we adopt spatio-temporal finite element simulations as introduced in \cite{pinon_measuring_2017,pinon_core-shell_2018}.
In these, an outer part of the sample containing the unpaired electrons is considered to be hyperpolarized to the steady-state polarization continuously with spin diffusion transporting this hyperpolarization towards the core of the particles with relaxation affecting each element at all times.
With $D$ from the above simulations, all parameters except the relaxation times $T_\mathrm{1,in}$ and $T_\mathrm{1,out}$ in the inner (core) and outer (shell) part of the particles are fixed.
Simulated build-ups and the best-fit parameters are given in Fig.~\ref{fig:Fig6}a.
For completeness, both $D$ from the nearest neighbor approach (Eqs.~\eqref{eq:nearestNeighbor}, $D=3.6$\,nm\textsuperscript{2}/s) and the lattice model (Eq.~\eqref{eq:D_lattice}, $D=51$\,nm\textsuperscript{2}/s) were used in the finite element simulations providing similar relaxation times at 7\,T.
For the 3.4\,T simulations, the fewer and lower SNR data points complicate the simulations as evident by the weaker variation of the calculated residues with variation of the relaxation times shown in Fig.~S12 of the Supplementary Material \cite{US10nmSM}.
Nevertheless, at both magnetic fields the general trend of $T_\mathrm{1,out}$ similar to the build-up time and much longer $T_\mathrm{1,in}$ persists.  

The long $T_\mathrm{1,in}$ can be explained by the (assumed) relative absence of paramagnetic defects in the core (cf. Sec.~S3 of the Supplementary Material \cite{US10nmSM}, for a basic electron paramagnetic resonance (EPR) characterization of the sample).
With paramagnetic relaxation nearly absent, a rigid crystalline lattice with a Debye temperature of 645\,K and liquid-helium temperatures, the relaxation in the core would be vanishing as found in the simulations.
The relaxation in the outer (shell) part of the sample is attributed to the naturally forming paramagnetic defects (often called $P_b$ centers) between the silicon core and the oxide shell \cite{lee_decay_2011,von_witte_controlled_2024}.
We would like to note again, that we model the relaxation in the outer shell homogeneously while in reality this will depend on the distance to the nearest paramagnetic center.

The build-up simulations can be adjusted to simulate the measured room temperature decays (cf. Fig.~\ref{fig:Fig2}b) and are shown for the 20 and 50\,nm particles in Fig.~\ref{fig:Fig6}b and c.
The relaxation time in the inner (core) part of the particles is between 1 and 3.4\,h while at the outer (shell) part between 0.2 and 0.5\,h for both particles and both choices of spin diffusion coefficients (3.6 and 51\,nm\textsuperscript{2}/s).
For the faster spin diffusion, the simulation model shows a weaker sensitivity to $T_\mathrm{1,in}$ compared to slower spin diffusion (cf. Fig.~S13 of the Supplementary Material \cite{US10nmSM}). 

Both, the build-up as well as the decay simulations estimate the relaxation time in the core ($T_\mathrm{1,in}$) to be longer than the measured experimental time scale (build-up or decay time) while the relaxation time in the shell ($T_\mathrm{1,out}$) is similar to the experimental time scale.
Furthermore, the effect of faster spin diffusion reduces the model's sensitivity to $T_\mathrm{1,in}$.
With the relaxation in the shell determining the build-up or decay time, this results in a mono-exponential behavior as only one parameter is rate limiting \cite{von_witte_modelling_2023}.
Thus, the problem of understanding the polarization dynamics in silicon nanoparticles is mostly reduced to understanding the processes in the shell.
Below, we discuss the possible origin of the long relaxation times in silicon nanoparticles.

Owing to the low temperatures, the rigid silicon lattice and the observed dependence of the hyperpolarization dynamics (build-up and decay times) on the thermal electron polarization (temperature and field) as shown in Figs.~\ref{fig:Fig2}a and~S7 of the Supplementary Material \cite{US10nmSM}, as well as Ref.~\cite{von_witte_controlled_2024}, the nuclear relaxation is attributed to paramagnetic relaxation.
The measured EPR spectra indicate a mixture of different $P_b$ defects (cf. Sec.~S3  of the Supplementary Material \cite{US10nmSM}).
The $P_b$ center is described as a dangling bond centered on a silicon atom with a 200-400\,MHz hyperfine coupling to the central Si site, $\sim 40$\,MHz for next nearest Si atoms and around 10\,kHz for Si atoms two unit cells away \cite{von_witte_controlled_2024}.
With the paramagnetic relaxation rate and the DNP transfer rate depending on the square of the hyperfine coupling, the central Si atom of the $P_b$ center will hyperpolarize and relax orders of magnitude faster than the other Si sites close to a $P_b$ center if it is occupied with a NMR-active \textsuperscript{29}Si atom.
The fast hyperpolarization and decay of the central \textsuperscript{29}Si of a $P_b$ center suggests that these sites are the primary polarization pathway if the hyperpolarization could be transported to the bulk \cite{von_witte_controlled_2024}.

The strong frequency differences between nuclei near the $P_b$ center due to different hyperfine couplings suppresses spin diffusion by nuclear dipolar flip-flops.
However, the electron itself modifies the nuclear spin diffusion in its vicinity, enabling non-energy conserving flip-flops.
Specifically, a term which describes an energy-conserving simultaneous electron flip-flop and nuclear flip-flop (electron-nuclear four-spin flip-flop) \cite{von_witte_two-electron_2024,redrouthu_overcoming_2024}, which is proportional to the electron dipolar coupling $D_\mathrm{ee}$ (mediating electron flip-flops) and the nuclear dipolar coupling $d_\mathrm{nn}$ (nuclear flip-flops) could cause nuclear spin diffusion close to the electron.
Isotope enrichment, i.e. the increase in \textsuperscript{29}Si abundance, would increase the nuclear dipolar couplings, eventually resulting in an increased electron-nuclear four-spin flip-flop rate.

In Ref.~\cite{kim_29si_2021}, silicon nanoparticles were synthesized with 4.7\% (natural abundance), 10\% and 15\% \textsuperscript{29}Si abundance.
After hyperpolarization, the room temperature $T_1$ times of the different particles were measured with 20\textdegree\ monitoring RF pulses every 4\,min.
The measured $T_1$ times were 48, 35 and 17\,min for 4.7\%, 10\% and 15\% \textsuperscript{29}Si, respectively \cite{kim_29si_2021}.
After correcting for the monitoring RF pulses \cite{von_witte_modelling_2023}, the relative ratio for relaxation rates ($1/T_1$) between the 4.7\%, 10\% and 15\% \textsuperscript{29}Si particles is 1:2.5:8.2.

The nuclear dipolar couplings $d_\mathrm{nn}$ scale with the third inverse power of the average nuclear-nuclear distance ($d_\mathrm{nn} \propto r_\mathrm{n-n}^{-3}$).
With $r_\mathrm{n-n}^{-3}$ inversely proportional to the isotope abundance $f$, the dipolar coupling is proportional to the isotope abundance ($d_\mathrm{nn} \propto f$). 
Assuming a quadratic dependence on the electron-nuclear four-spin flip-flop effective Hamiltonian matrix element, the nuclear flip-flop rate (spin diffusion) close to the electron would scale with the dipolar coupling squared and, hence, with the isotope abundance squared.
The relative ratio of the isotope abundances squared between the 4.7\%, 10\% and 15\% \textsuperscript{29}Si particles is 1:4.5:10.2.
Considering this simple estimation, the ratios between the isotope abundances are compatible with the RF corrected relaxation rates.
Remaining discrepancies might in parts arise from effects not considered in the simple estimation above, e.g. reduced frequency differences between neighboring \textsuperscript{29}Si atoms near the electrons due to their smaller distance and higher numbers of \textsuperscript{29}Si atoms to couple to. 
For reduced frequency differences, the electron line shape might cause a further increase of the electron-nuclear four-spin flip-flop rate.

We highlight that an increase in the \textsuperscript{29}Si abundance leaves the total number of \textsuperscript{29}Si nuclei per $P_b$ center with a \textsuperscript{29}Si atom at its strongly hyperfine coupled central site constant.
Thus, it appears unlikely that the long room temperature relaxation times of silicon nanoparticles are simply due to inefficient paramagnetic relaxation by the $P_b$ centers as in this case the $T_1$ relaxation times should be rather independent of the isotope abundance.

\subsection*{Generalization to other crystal structures}

In the following, we will adopt the SQ and ZQ line width as well as the spin-diffusion simulations to simple cubic, body-centered cubic (BCC), face-centered cubic (FCC) and diamond cubic crystal structures.
The gyromagnetic ratio is set to $10^6~\mathrm{rad/(s \cdot T)})$ and the lattice constant to 1\,\AA.
An effective formula to estimate both the line width and spin diffusion is given based on the simulations. 
Therefore, adoption to other spin-1/2 crystals involves the change of gyromagnetic ratio, lattice constant and isotope abundance as discussed below.
We note that for crystals with a multi-atomic basis the spin diffusion of one species can be described by the sub-lattice of the respective atomic species e.g. in sodium fluoride (NaF) each of the two atomic species forms a FCC lattice although heteronuclear dipolar couplings might influence the line widths.

\begin{figure}[ht]
	\centering
	\includegraphics[width=\linewidth]{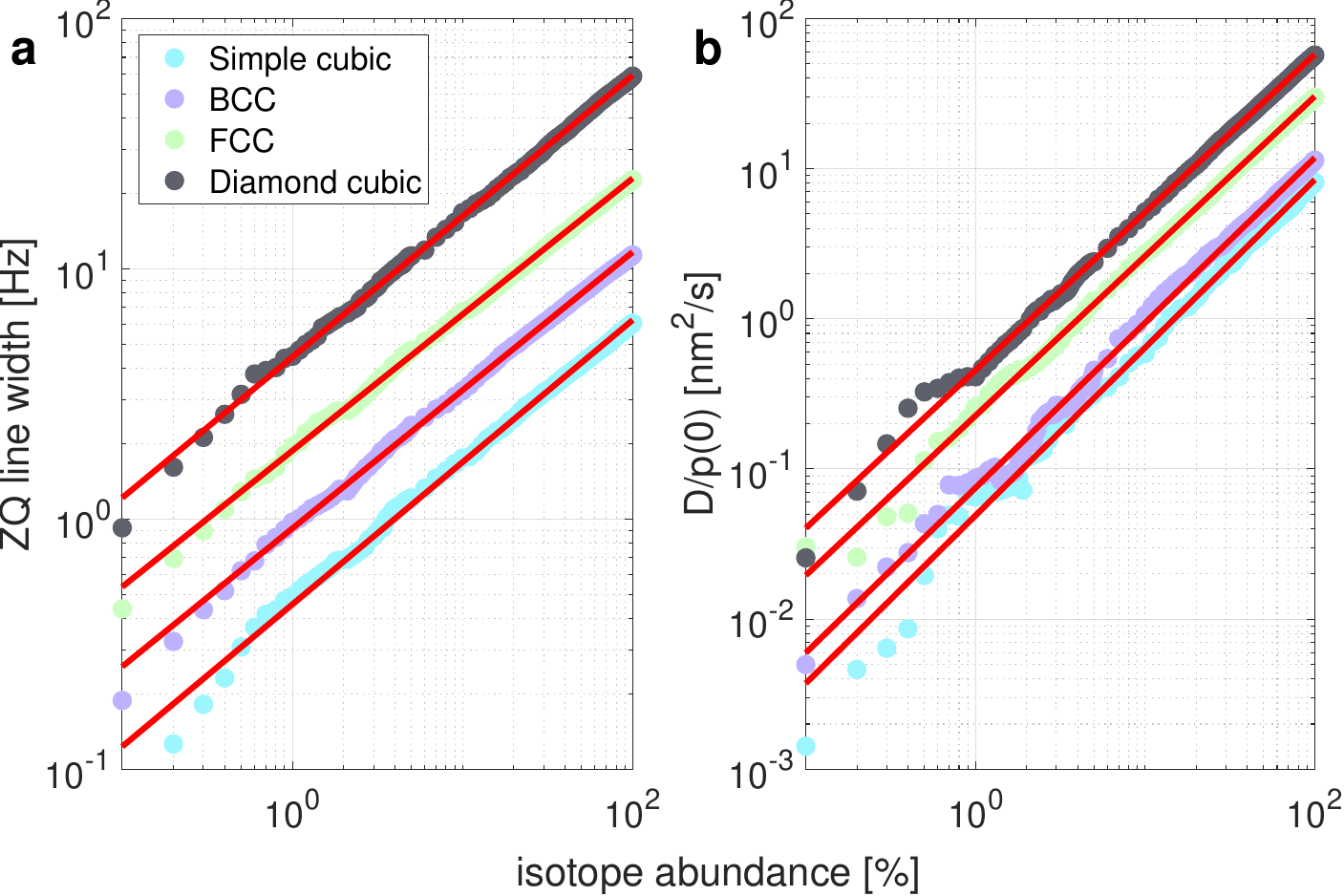}
	\caption{\textbf{(a)} Simulated ZQ line width and \textbf{(b)} geometric contribution to the spin diffusion coefficient $D/p(0)$ for generalized simple cubic, body-centered cubic (BCC), face-centered cubic (FCC) and diamond cubic lattices.
    The fit parameters are summarized in Tab.~\ref{tab:fitParameter_ZQ_D} and can be used to calculate the respective quantities with an analytic expression (cf. main text and Eqs.~\eqref{eq:ZQlineWidthGeneral} and~\eqref{eq:D_general}).}
	\label{fig:Fig7}
\end{figure}

In Sec.~S7 of the Supplementary Material \cite{US10nmSM}, the dipolar cut-off distance similar to Fig.~\ref{fig:Fig3}b and the SQ and ZQ line widths similar to Fig.~\ref{fig:Fig4}a are compared for all considered crystal structures.
The results are very similar to the above presented case of silicon.
The ZQ lines of the different crystals structures are compared in Fig.~\ref{fig:Fig7}a in a double logarithmic plot. 
The increasing number of atoms per conventional unit cell (1, 2, 4, 8 for simple cubic, BCC, FCC and diamond cubic) results in stronger dipolar couplings and with this larger line widths.

Experimentally, the ZQ line width is difficult to measure.
Fortunately, the SQ line width is a good approximation for the ZQ line width (cf. Fig.~S11 of the Supplementary Material \cite{US10nmSM}) and this can be experimentally measured.
The spectral density $p$ of the fitted experimental line shape evaluated for zero frequency ($p(0)$) can be multiplied with prefactors described in Eq.~\eqref{eq:D_lattice}.
Fig.~\ref{fig:Fig7}b shows the spin diffusion coefficient $D$ divided by $p(0)$, which is a geometric quantity with some prefactors.

The fit parameters corresponding to Fig.~\ref{fig:Fig7} for the ZQ line widths and $D/p(0)$ are summarized in Tab.~\ref{tab:fitParameter_ZQ_D}.
Multiplication with the isotope abundance, gyromagnetic ratio and lattice constant gives rise to the ZQ (and to some degree SQ) line width and spin diffusion coefficient.
Specifically,
\begin{align}
    \Delta \nu_\mathrm{ZQ} &= \frac{\Tilde{\gamma}_n^2 }{\Tilde{a}_0^3} u_\mathrm{ZQ} f^{m_\mathrm{ZQ}} \label{eq:ZQlineWidthGeneral}\\
    D &= \frac{\Tilde{\gamma}_n^2 }{\Tilde{a}_0^2} u_D f^{m_D} p(0) \label{eq:D_general}
\end{align}
with $\Tilde{\gamma}_n = \gamma_n/(10^6~\mathrm{rad/(s \cdot T)})$, $\Tilde{a} = a/(1~\mathrm{\AA})$ the rescaled nuclear gyromagnetic ratio and lattice constant.
$\gamma_n$ is the gyromagnetic ratio, e.g. $53.190 \cdot 10^6$\,rad/(s $\cdot$ T) for \textsuperscript{29}Si.
$f$ is the isotope abundance in \%. 
$u$ and $m$ are summarized in Tab.~\ref{tab:fitParameter_ZQ_D} and describe the dependence of the spin diffusion on the underlying crystalline lattice and isotope abundance.
To give an example of Eq.~\eqref{eq:ZQlineWidthGeneral}, for 4.7\% \textsuperscript{29}Si abundance, $\Tilde{\gamma}_n = 53.190$, $\Tilde{a} = 5.431$, $u_\mathrm{ZQ} = 4.44$ and $m_\mathrm{ZQ} = 0.563$, we find a ZQ line width of 187\,Hz in good agreement with the simulated 191\,Hz from Fig.~\ref{fig:Fig4} (cf. discussion of this value above).

\begin{table}[htbp]
\centering
\caption{Fit parameters from Fig.~\ref{fig:Fig7}, fitted with Eqs.~\eqref{eq:ZQlineWidthGeneral} and~\eqref{eq:D_general}. }
\begin{tabular}{l|l|l|l|l}
 & simple cubic & BCC & FCC & diamond cubic \\ \hline
$u_\mathrm{ZQ}$ & 0.456(11) & 0.918(12) & 1.88(2) & 4.44(5) \\ 
$m_\mathrm{ZQ}$ & 0.568(8) & 0.552(4) & 0.544(3) & 0.563(4) \\ \hline
$u_\mathrm{D}$ & 0.049(3) & 0.075(3) & 0.226(6) & 0.455(11) \\
$m_\mathrm{D}$ & 1.118(17) & 1.099(10) & 1.063(9) & 1.052(8) \\ 
\end{tabular}
\label{tab:fitParameter_ZQ_D}
\end{table}

The above approach to generate effective scalings for the ZQ/ SQ line widths and spin diffusion coefficient could be used to derive similar expressions for other crystal structures. 
The used simulation files can be found online (cf. Materials \& Correspondence section).


\section*{Conclusions}

Nanoparticles a few tens of nm in size can exhibit hyperpolarization enhancements and relaxation times similar to larger \textmu m-sized particles if the surface defects (dangling bonds) are inhibited from causing a strong paramagnetic relaxation.

Lattice simulations of the single- and zero-quantum lines can be used to simulate the build-up and decay dynamics of the particles, which reveal that the polarization dynamics is determined by the shell (outer) part of the particles containing the interface defects.



\nocite{stesmans_electron_1997,van_gorp_dipolar_1992,cheng_investigations_1973}

\begin{acknowledgments}
We thank Frank Krumeich for collecting TEM images and  ScopeM, ETH Zurich, for access to the imaging facilities.

ME acknowledges support by the Schweizerischer Nationalfonds zur Förderung der Wissenschaftlichen Forschung (grant no. 200020\_188988 and 200020\_219375).
KT and JOM acknowledge support by Research Council of Finland (grant no. 331371, 338733, and Flagship of Advanced Mathematics for Sensing Imaging and Modelling grant 358944), Finnish Cultural Foundation (North Savo regional fund) and Saastamoinen Foundation.
Financial support of the Horizon 2020 FETFLAG MetaboliQs grant is gratefully acknowledged. \\
\end{acknowledgments}

\section*{Author Contributions Statement}
GvW, GK, SK and ME conceptualized the research. 
GvW, KT, N\c{S}, AH, VG and GK performed experiments and analyzed the data. 
GvW and ME developed the simulation model.
GvW and N\c{S} implemented and performed the simulations. 
GvW prepared the original draft. 
SK and ME acquired funding and provided supervision. 
JOM, VPL SK and ME provided resources.
All authors reviewed and edited the draft.

\section*{Competing Interests Statement}
The authors declare that they have no competing interests. 

\section*{Materials \& Correspondence}
Experimental and simulation data together with Matlab scripts can be found under \url{https://doi.org/10.3929/ethz-b-000709810}.
Further correspondence should be addressed to Matthias Ernst (maer@ethz.ch).

\bibliography{references,notes}

\end{document}


\renewcommand{\figurename}{Fig.}


\title{Supplementary Material: The role of nuclear spin diffusion in dynamic nuclear polarization of crystalline nanoscale silicon particles}

\author{Gevin von Witte}
 \affiliation{Institute for Biomedical Engineering, University and ETH Zurich, 8092 Zurich, Switzerland}
 \affiliation{Institute of Molecular Physical Science, ETH Zurich, 8093 Zurich, Switzerland}
\author{Konstantin Tamarov}
 \affiliation{Department of Technical Physics, University of Eastern Finland, 70210 Kuopio, Finland}
\author{Neva \c{S}ahin}
 \affiliation{Bilkent University, Ankara, Turkey}
\author{Aaron Himmler}
 \affiliation{Institute of Molecular Physical Science, ETH Zurich, 8093 Zurich, Switzerland}
\author{Vera Ganz}
 \affiliation{Institute of Molecular Physical Science, ETH Zurich, 8093 Zurich, Switzerland}
\author{Jani O. Moilanen}
 \affiliation{Department of Chemistry, Nanoscience Center, 
University of Jyväskylä, 40500 Jyväskylä, Finland}
\author{Vesa-Pekka Lehto}
 \affiliation{Department of Technical Physics, University of Eastern Finland, 70210 Kuopio, Finland}
\author{Grzegorz Kwiatkowski}
 \affiliation{Institute for Biomedical Engineering, University and ETH Zurich, 8092 Zurich, Switzerland}
\author{Sebastian Kozerke}
 \affiliation{Institute for Biomedical Engineering, University and ETH Zurich, 8092 Zurich, Switzerland}
\author{Matthias Ernst}
 \affiliation{Institute of Molecular Physical Science, ETH Zurich, 8093 Zurich, Switzerland}
 \email{maer@ethz.ch}

\date{\today}

\maketitle

\onecolumngrid

\setcounter{equation}{0}
\setcounter{figure}{0}
\setcounter{table}{0}
\setcounter{section}{0}
\setcounter{page}{1}
\makeatletter
\renewcommand{\theequation}{S\arabic{equation}}
\renewcommand{\thefigure}{S\arabic{figure}}
\renewcommand{\thetable}{S\arabic{table}}
\renewcommand{\thesection}{S\arabic{section}}

\tableofcontents

\clearpage

\section{Average particle size and X-ray powder diffraction}

\begin{figure}[ht]
	\centering
	\includegraphics[width=0.5\linewidth]{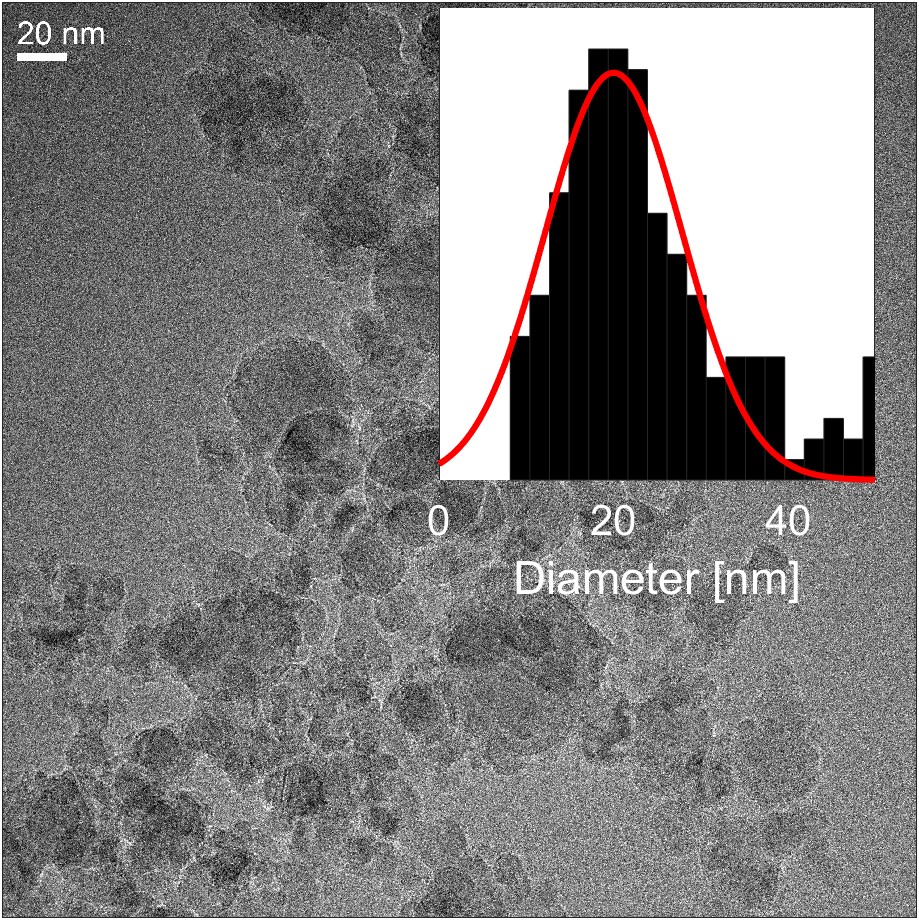}
	\caption{TEM image of the particles and particle size distribution (inset, average particle size (APS) 20$\pm$12\,nm).}
	\label{fig:SI_APS}
\end{figure}

\begin{figure}[ht]
    \centering
    \includegraphics[width=\linewidth]{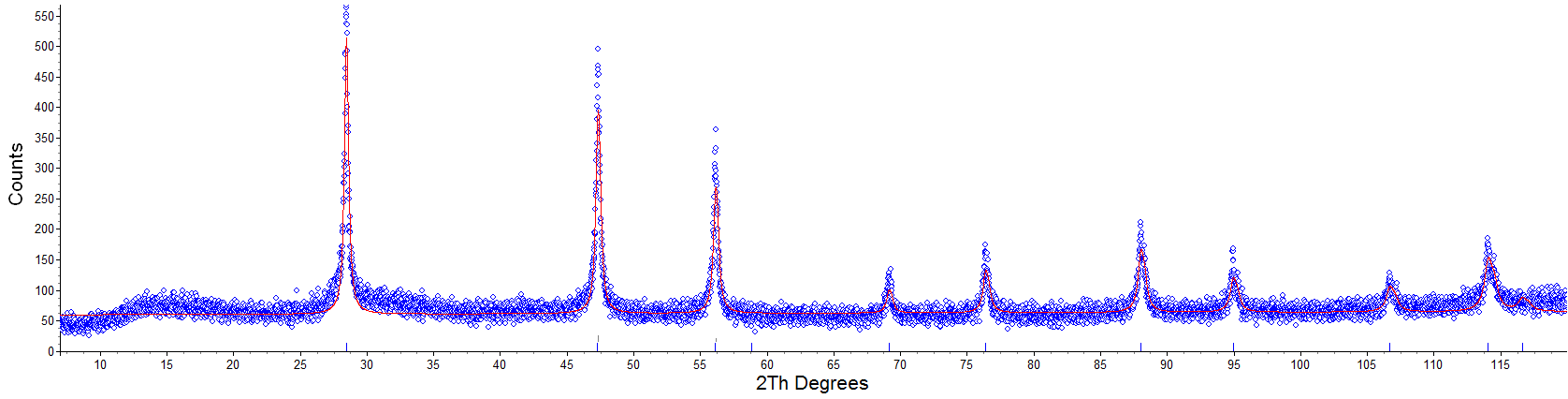}
    \caption{X-ray powder diffraction patter of the particles (blue circles) and its fitting according to the Rietvield refinement method (red line).}
    \label{fig:SI_XRPD}
\end{figure}

\clearpage

\section{DNP profiles} \label{sec:SI_Profiles}

\begin{figure*}[ht]
	\centering
	\includegraphics[width=\linewidth]{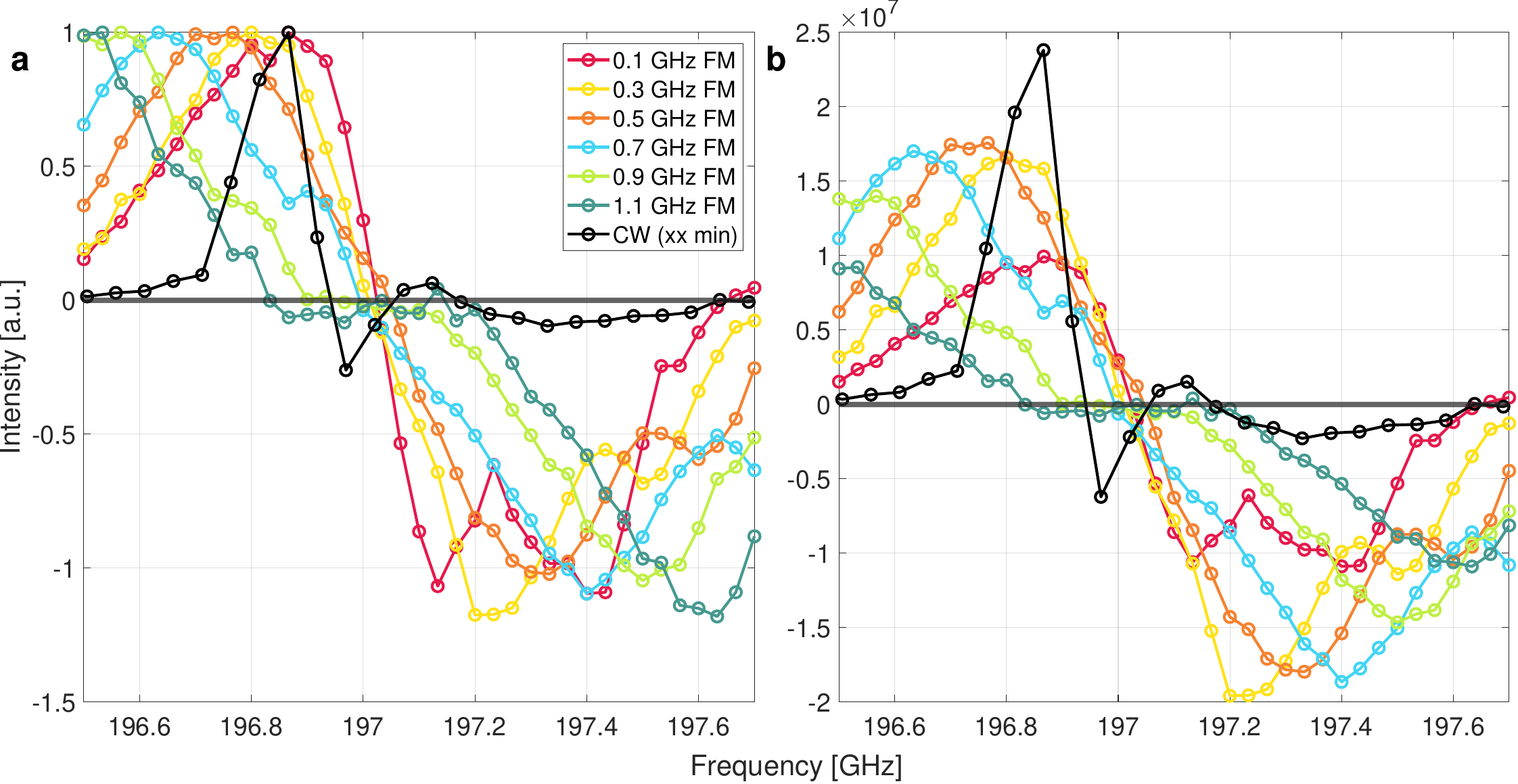}
	\caption{\textbf{(a)} Normalized and \textbf{(b)} non-normalized DNP profiles of the 20\,nm particles recorded at 7\,T with a 200\,mW VDI source and low-loss electroplated wave guides \cite{himmler_electroplated_2022}, nearly doubling the MW power reaching the sample. 
    The CW DNP profile shows nearly no negative DNP enhancement (tested for different MW irradiation times, data not shown).
    Upon the addition of FM, the profile becomes nearly symmetric.}
	\label{fig:SI_profile20nm7T}
\end{figure*}

\begin{figure*}[ht]
	\centering
	\includegraphics[width=0.7\linewidth]{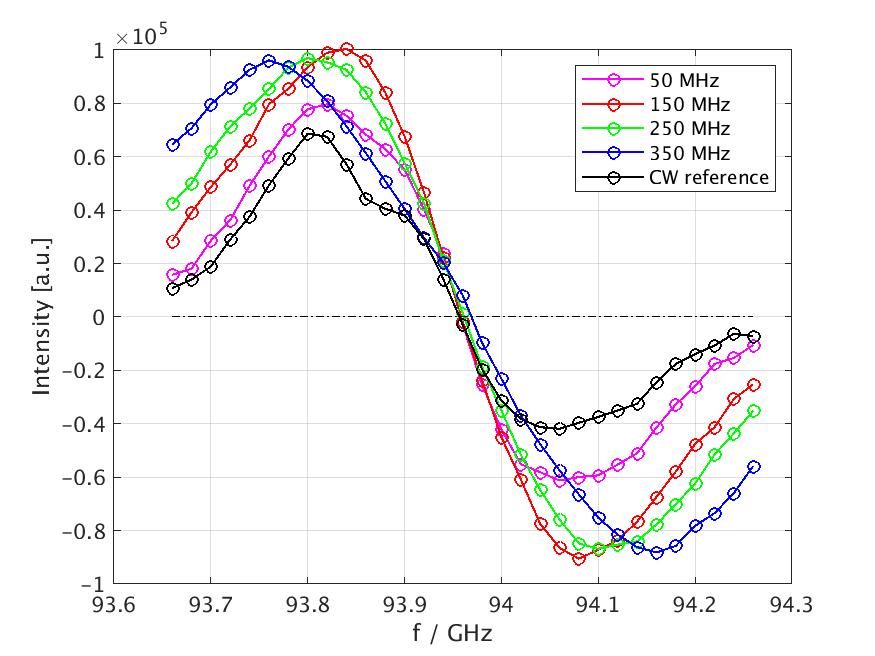}
	\caption{DNP profile of the 20\,nm particles recorded at 3.4\,T with a more powerful 400\,mW VDI source. 
    The CW DNP profile shows a reduced negative DNP enhancement compared to the positive DNP enhancement lobe.
    Upon the addition of FM, the profile becomes nearly symmetric.
    A larger FM bandwidth stretches the DNP profile but does not increase enhancements compared to a 150\,MHz modulation.
    The peak-to-peak frequency difference for the 150\,MHz modulation is around 240\,MHz ((\textsuperscript{29}Si Larmor frequency around 28\,MHz), similar to the (CW) DNP profile of the 50\,nm particles (cf. Fig.~\ref{fig:SI_profile50nm3T}).}
	\label{fig:SI_profile20nm3T}
\end{figure*}

Frequency modulation (FM) of the microwave (MW) irradiation was previously shown to improve the DNP of silicon particles \cite{kwiatkowski_nanometer_2017,kwiatkowski_exploiting_2018}.
FM modulation stretches the DNP profile, which we observed for the 20\,nm particles in Figs.~2c of the main manuscript, \ref{fig:SI_profile20nm7T} and \ref{fig:SI_profile20nm3T}.
For continuous wave (CW) irradiation, without FM, the DNP profiles of the 20\,nm particles is asymmetric between the positive and negative DNP enhancement lobes as shown in Figs.~2c of the main manuscript, \ref{fig:SI_profile20nm7T} and \ref{fig:SI_profile20nm3T}.
The origin of this asymmetry remains unclear.
Upon the addition of FM, the profiles become much more symmetric.

\begin{figure*}[ht]
	\centering
	\includegraphics[width=0.6\linewidth]{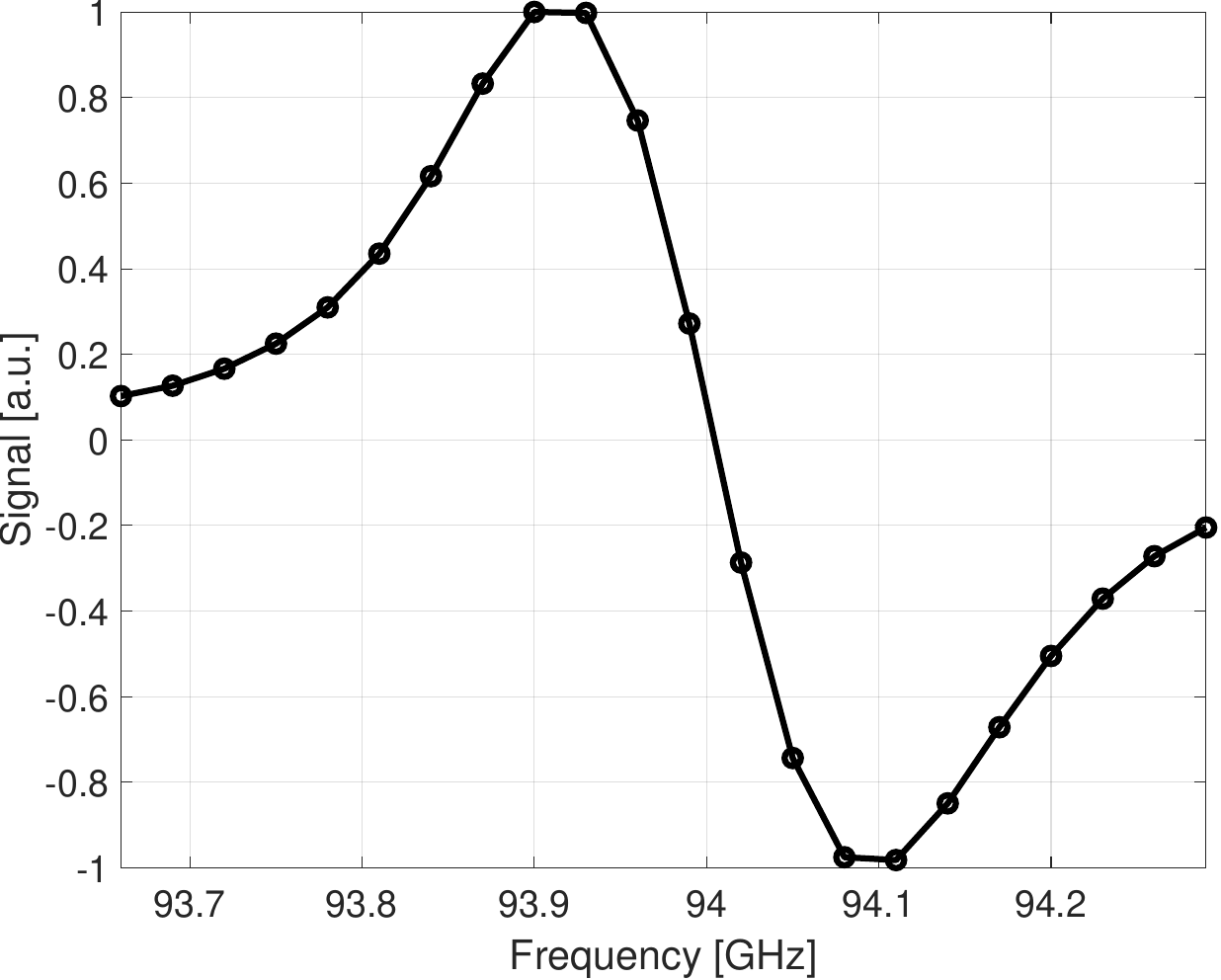}
	\caption{DNP profile of the 50\,nm particles recorded at 3.4\,T with continuous wave (CW) MW irradiation as previously published \cite{kwiatkowski_exploiting_2018}. 
    The peak-to-peak frequency difference is around 200\,MHz (\textsuperscript{29}Si Larmor frequency around 28\,MHz). 
    This profile was recorded before a magnet quench with the field slightly different than for the other measurements shown in this work.}
	\label{fig:SI_profile50nm3T}
\end{figure*}

With a 150\,MHz bandwidth FM at 3.4\,T, the DNP profile of the 20\,nm sample looks similar to the (CW) profile of the 50\,nm particles (cf. Figs.~\ref{fig:SI_profile20nm3T} and~\ref{fig:SI_profile50nm3T}).
We note that the 50\,nm particles' profile has the same shape as for \textmu-sized particles \cite{kwiatkowski_exploiting_2018} and porous silicon (PSi) nanoparticles \cite{von_witte_controlled_2024}.

\clearpage

\section{Electron paramagnetic resonance} \label{sec:SI_EPR}

\begin{figure*}[ht]
	\centering
	\includegraphics[width=\linewidth]{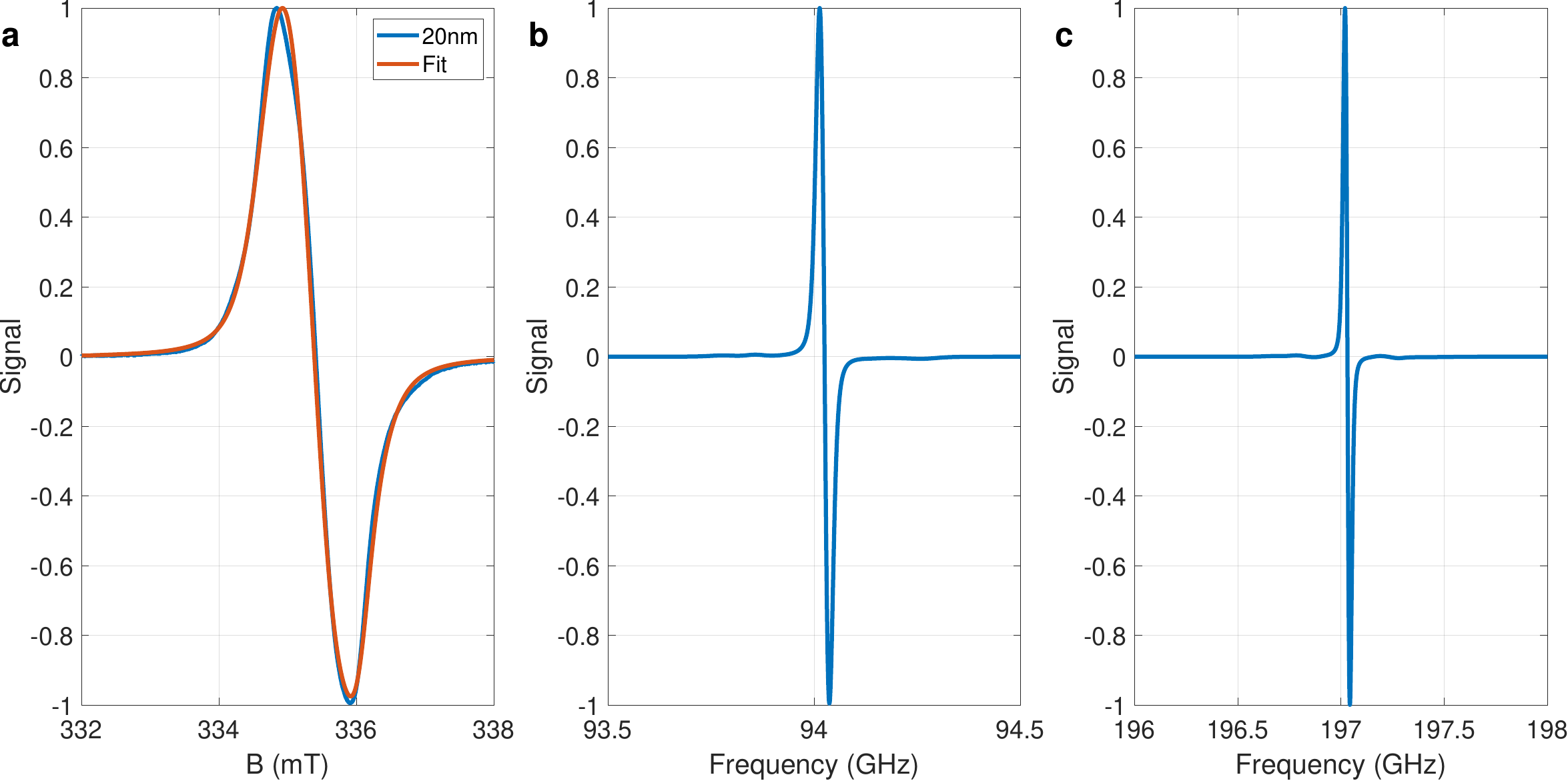}
	\caption{\textbf{(a)} Fit of the measured X-band EPR profile of the 20\,nm particles. The fit assumes the presence of $P_b^\mathrm{(111)}$ and isotropic $P_b^\mathrm{iso}$ centers \cite{von_witte_controlled_2024}.
    A density of $2.7\cdot 10^{15}$\,mg\textsuperscript{-1} $P_b$ centers was estimated with respect to a TEMPO sample of which approximately 19\% are $P_b^\mathrm{(111)}$ and 81\% $P_b^\mathrm{iso}$ centers.
    Extrapolation of the X-band EPR fit to (\textbf{b}) 3.35\,T and (\textbf{c}) 7.02\,T. }
	\label{fig:SI_EPRfit}
\end{figure*}

An X-band electron paramagnetic resonance (EPR) spectrum and its fitting with EasySpin are shown in Fig.~\ref{fig:SI_EPRfit}a.
The fitting follows the approach from \cite{von_witte_controlled_2024}.
Briefly, EasySpin fitted the powder average spectrum using the combination of anisotropic $P_b^\mathrm{(111)}$ and isotropic $P_b^\mathrm{iso}$ centers.
Anisotropic centers had the trigonal symmetry with $g_\parallel = 2.00185,\,A_\parallel = 230 \pm 25$\,MHz and $g_\perp = 2.0081,\,A_\perp = 420 \pm 15$\,MHz.
$g$-strain \cite{stesmans_electron_1997} and Lorentzian line broadening \cite{van_gorp_dipolar_1992} were included to account for the $g$-factor strain and $P_b$ dipolar interaction of $P_b^\mathrm{(111)}$ centers, respectively. $P_b^\mathrm{iso}$ centers were fitted with an isotropic $g$-factor (2.0053) a phenomenological Voigtian lineshape to include homogeneous and inhomogeneous line broadening effects due to strain and dipolar interaction \cite{von_witte_controlled_2024}.
For more details about the defects an an EPR summary of silicon nanoparticles, the reader is referred to \cite{von_witte_controlled_2024}.

The number of $P_b$ centers was calculated by double integration of the sample spectrum and the TEMPO spectrum with the known number of spins.
For the 20\,nm particles, we find a density of $(2.7 \pm 0.3)\cdot 10^{15}$\,mg\textsuperscript{-1} which is within a factor of two to the best porous silicon (PSi) nanoparticles \cite{von_witte_controlled_2024}.
From the EasySpin fitting, there are around 19\% are $P_b^\mathrm{(111)}$ centers and 81\% $P_b^\mathrm{iso}$ among the total number of centers.
The EasySpin fit is then used to simulate the EPR spectrum at the DNP conditions (Fig.~\ref{fig:SI_EPRfit}b and c).

\clearpage

\section{Parameters of the build-ups} \label{sec:ExperimentParameters}

\begin{figure*}[ht]
	\centering
	\includegraphics[width=\linewidth]{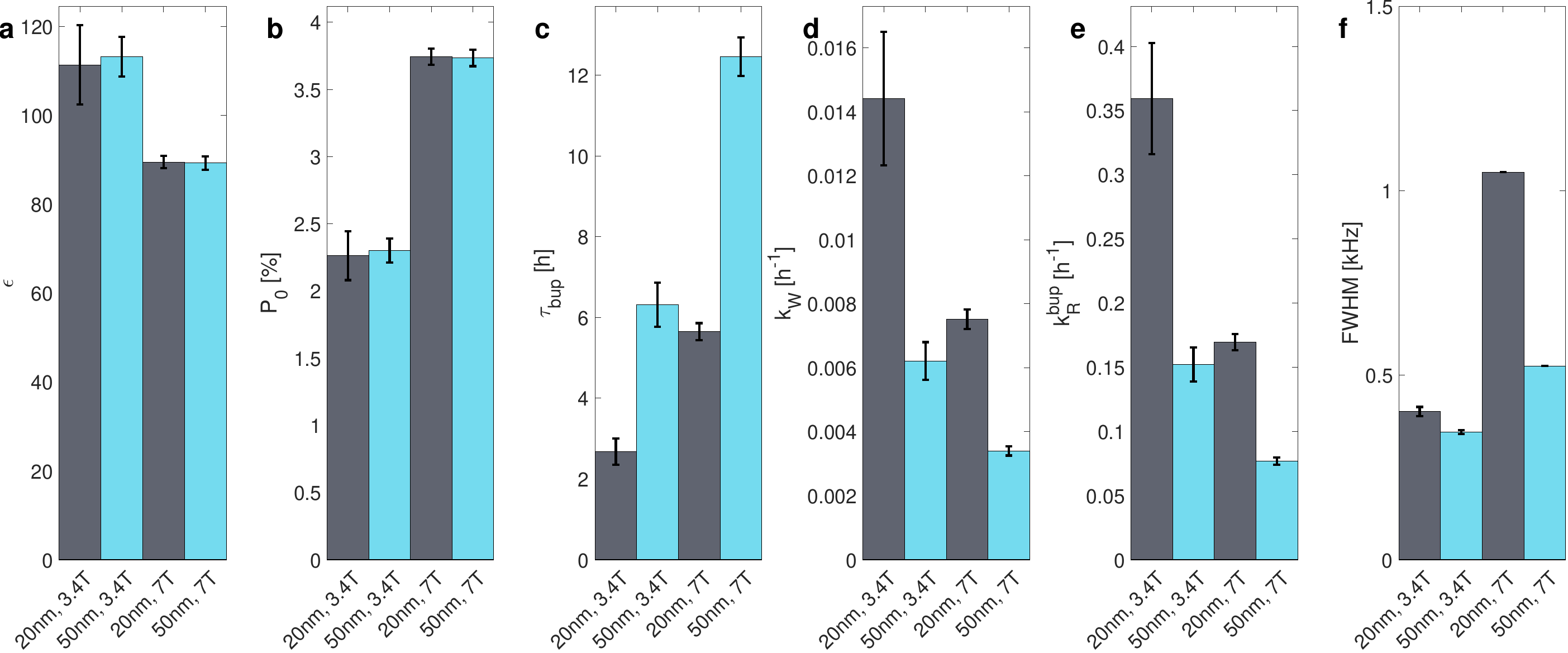}
	\caption{(\textbf{a}) Summary of the measured enhancements, (\textbf{b}) converted to polarization and (\textbf{c}) build-up time. 
    The experimental parameters can be converted to a (\textbf{d}) DNP injection rate and (\textbf{e}) relaxation rate during the build-up (cf. Eqs.~\eqref{eq:1compartment_bup_solution}).
    (\textbf{f}) The full width at half maximum (FWHM) of the Fourier transformed free induction decays (FIDs).
    See main text and text above for more details.}
	\label{fig:SI_experimentParameters}
\end{figure*}

Fig.~\ref{fig:SI_experimentParameters} shows the measured enhancement $\varepsilon$, steady-state nuclear polarization $P_0$ and build-up time $\tau_\mathrm{bup}$ for the 20 and 50\,nm samples at 3.4\,T and 7\,T.
The enhancements are slightly higher at 3.4\,T while the higher thermal nuclar polarization at 7\,T leads to a higher polarization. 
The 20\,nm particles build-up nearly twice as fast as the 50\,nm particles at 3.4\,T and 7\,T.
The build-ups at 7\,T take approximately twice as long as at 3.4\,T for both samples.

The steady-state polarization $P_0$ and build-up time $\tau_\mathrm{bup}$ can be used to calculate the DNP injection $k_\mathrm{W}$ and relaxation rate constants during the build-up $k_\mathrm{R}^\mathrm{bup}$ \cite{von_witte_modelling_2023,von_witte_relaxation_2024}:
\begin{align}
	\frac{\text{d}P}{\text{d}t} = (A-P) k_\mathrm{W} - k_\mathrm{R}^\mathrm{bup} P \label{eq:ODE_bup_1compartment}
\end{align}
for which the solution is given by
\begin{subequations} \label{eq:1compartment_bup_solution}
	\begin{align}
		\tau_{\mathrm{bup}}^{-1} &= k_\mathrm{W}+k_\mathrm{R}^\mathrm{bup} \label{eq:1compartment_bup_tau} \\
		P_0 &= \frac{Ak_\mathrm{W}}{k_{W}+k_\mathrm{R}^\mathrm{bup}} = Ak_\mathrm{W}\tau_{\mathrm{bup}} \label{eq:1compartment_bup_P0}
	\end{align}
\end{subequations}
The calculated rate constants are given in Fig.~\ref{fig:SI_experimentParameters}d, e.
The fast build-up of the 20\,nm particles is achieved by high injection and relaxation rates, even compared to custom synthesized porous silicon (PSi) nanoparticles \cite{von_witte_controlled_2024}.

At 7\,T, we measured the relaxation of the 50\,nm particles at 3.4\,K and found a RF pulse corrected \cite{von_witte_modelling_2023} relaxation time around 28\,h, corresponding to a relaxation rate of 0.036\,h\textsuperscript{-1}.
Contrasting this with the relaxation rate $k_\mathrm{R}^\mathrm{bup}$ during the build-up of 0.077\,h\textsuperscript{-1}, this suggests a relaxation enhancement by MW irradiation \cite{von_witte_relaxation_2024,von_witte_controlled_2024} by around two and with this limiting the achievable polarization levels.

Fig.~\ref{fig:SI_experimentParameters}f compares the line width (full width at half maximum, FWHM) of the free induction decay (FID) between the samples and fields.
While at 3.4\,T the 20 and 50\,nm particles have nearly identical line widths.
At 7\,T, the line width of the 20\,nm sample more than doubles compared to 3.4\,T.
In contrast, the line width of the 50\,nm particles remains nearly unchanged upon changing the field.

\clearpage

\section{Signal during room temperature decay}

\begin{figure*}[ht]
	\centering
	\includegraphics[width=\linewidth]{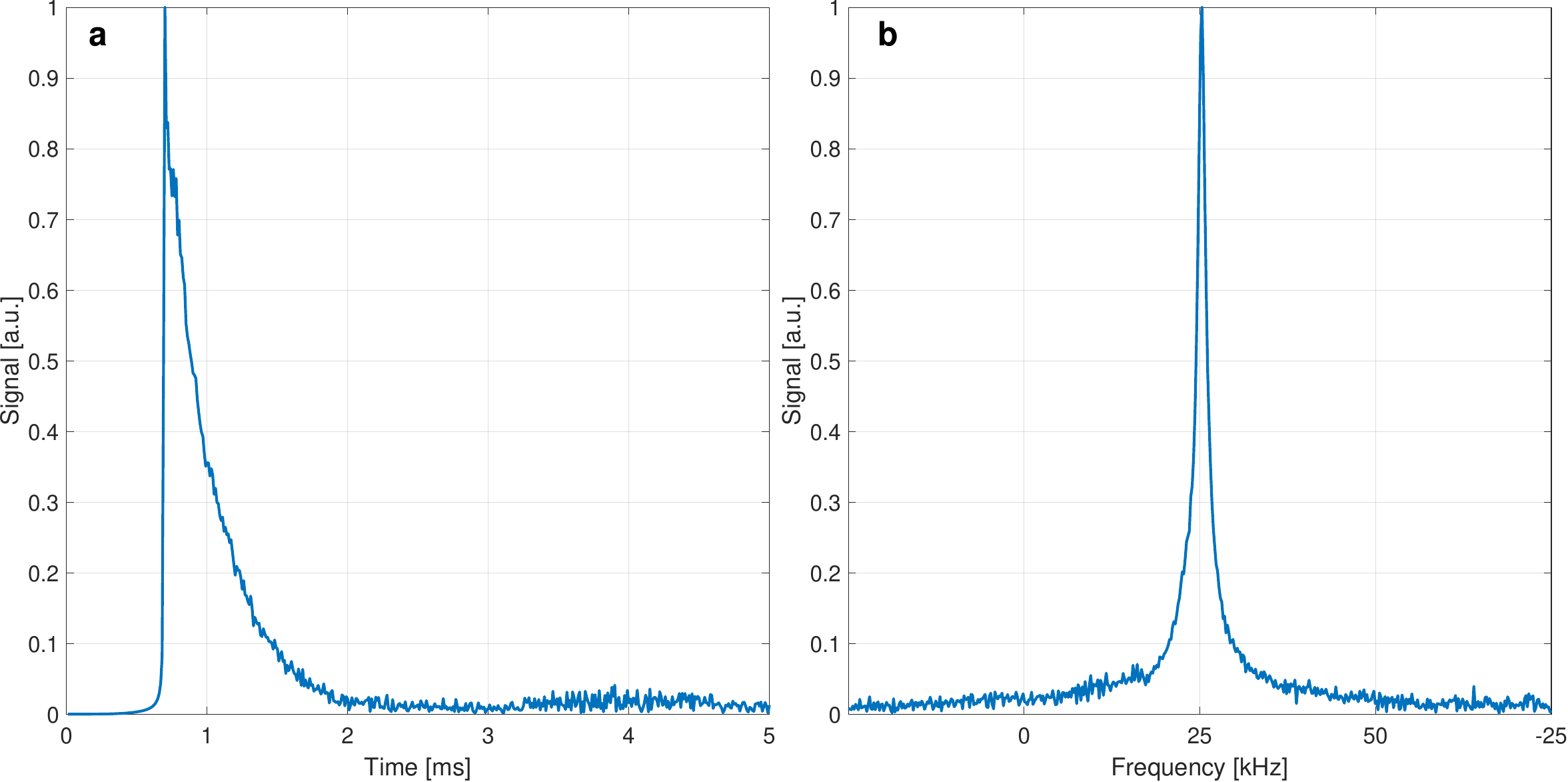}
	\caption{(\textbf{a}) Time domain and (\textbf{b}) frequency domain signal detected during the beginning of room temperature decay of the 20\,nm particles.}
	\label{fig:SI_AICsignal}
\end{figure*} 

\clearpage

\section{Simulated angular dependence of the single (SQ) and zero quantum (ZQ) lines}

\begin{figure*}[ht]
	\centering
	\includegraphics[width=0.7\linewidth]{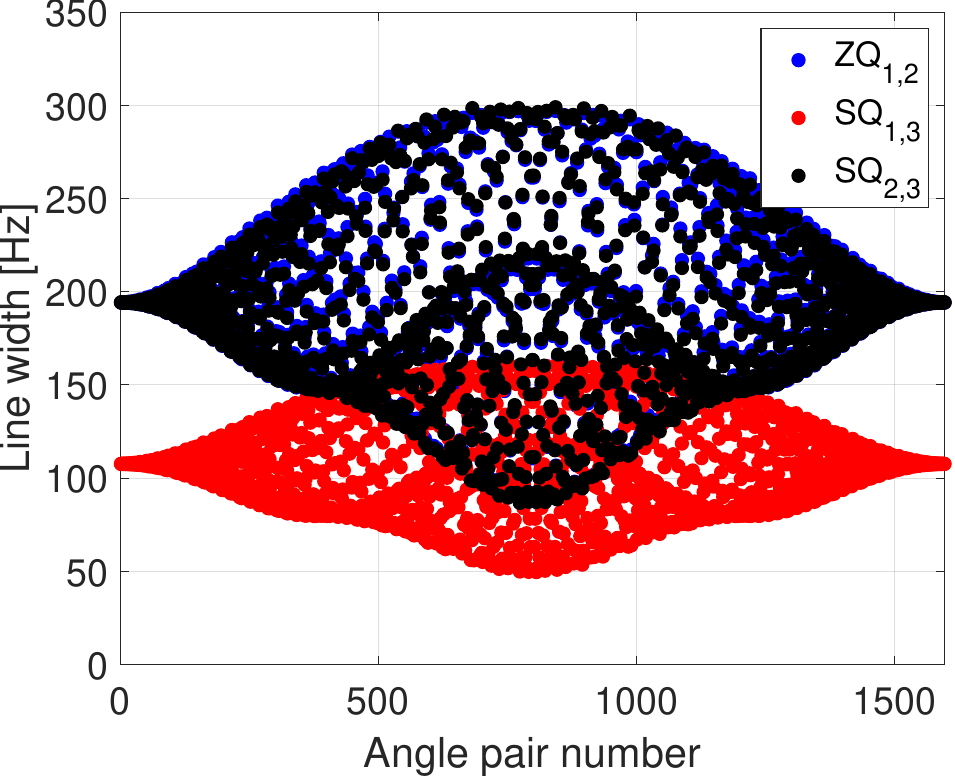}
	\caption{Angular dependence of the single-quantum (SQ) and zero-quantum (ZQ) line widths for a 4.7\% (natural) \textsuperscript{29}Si abundance structure after averaging over 100 random lattices.
    The individual lines vary by around $\pm$50\% of their average value depending on the spatial orientiation.
    Owing to the chosen averaging scheme, the line width is given in terms of $\alpha$-$\beta$ pairs which are ordered by an angle pair number.}
	\label{fig:SI_AngularDependence}
\end{figure*}

In the SQ, ZQ and spin diffusion simulations powder averaging was employed (1597 $\alpha$-$\beta$ pairs of Zaremba-Cheng-Wolfsberg (ZCW) averaging \cite{cheng_investigations_1973}) 
The SQ and ZQ lines show an angular dependence which is shown for natural abundance after averaging over 100 random lattices in Fig.~\ref{fig:SI_AngularDependence}.
The individual lines can vary by around $\pm$50\% around their average value depending on the spatial orientation.

\clearpage

\section{Generalization of spin diffusion simulations to other crystals} \label{sec:SI_generalSpinDiffusion}

\begin{figure*}[ht]
	\centering
	\includegraphics[width=0.6\linewidth]{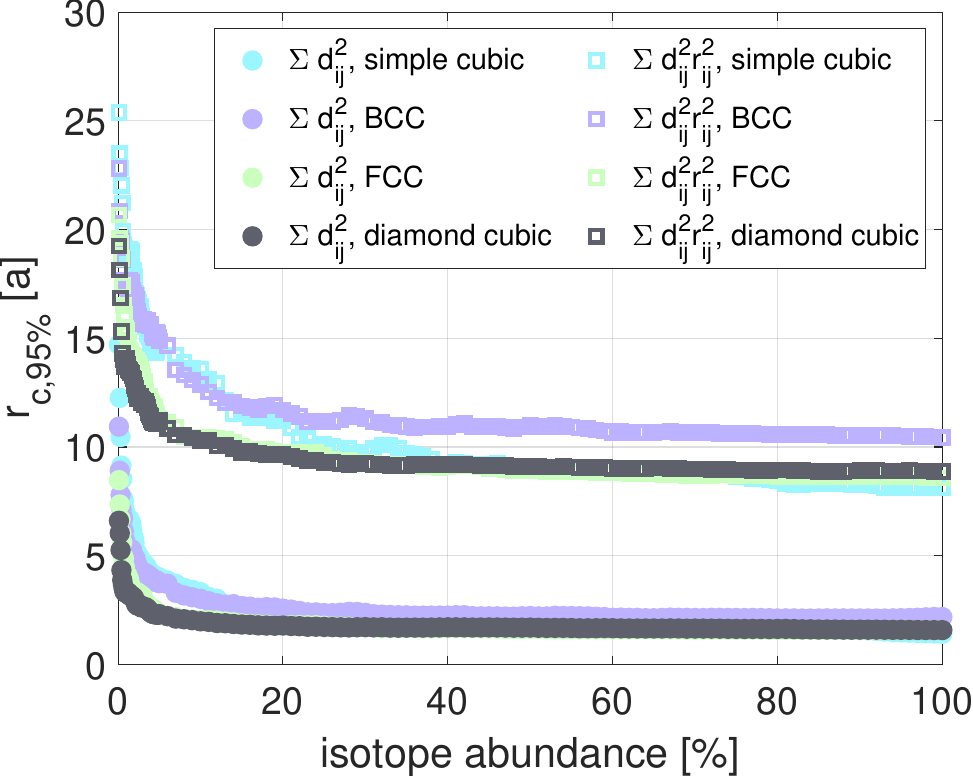}
	\caption{Cut-off for the squared dipoalr interaction ($d_{ij}^2$) as required for the SQ and ZQ simulations and geometric part of the spin diffusion coefficient ($d_{ij}^2$, cf. Eq.~6 of the main manuscript, without the geometrical dependence of the line width).}
	\label{fig:SI_dipolarCutoffGeneral}
\end{figure*}

\begin{figure*}[ht]
	\centering
	\includegraphics[width=\linewidth]{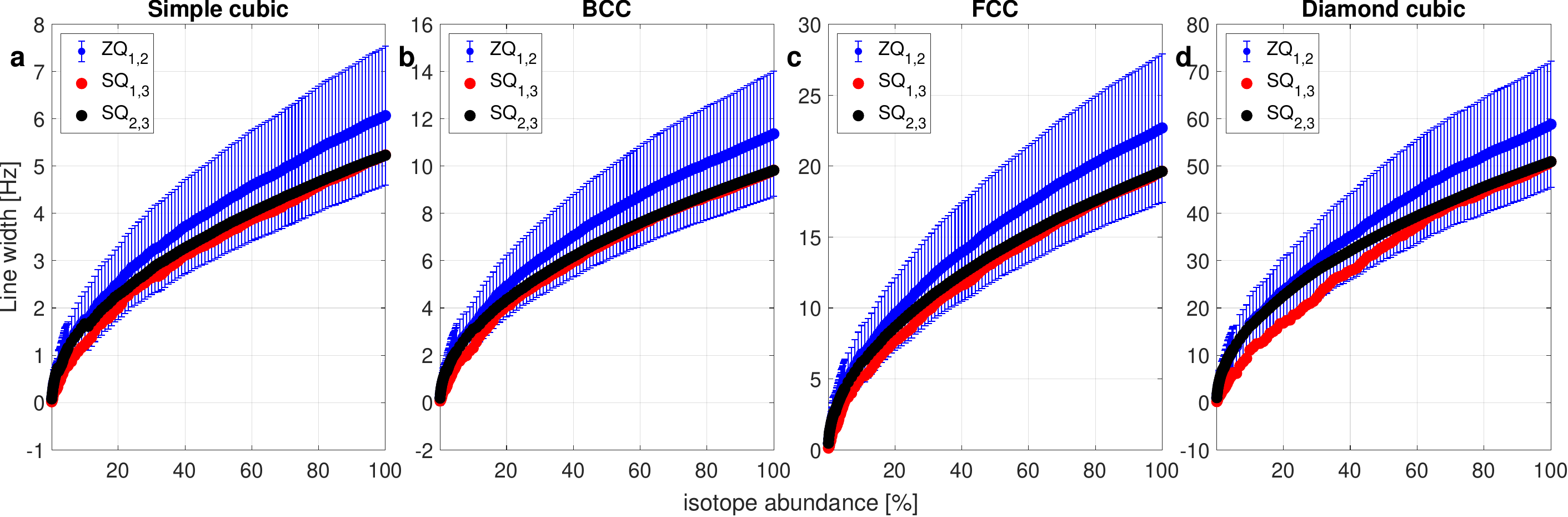}
	\caption{Comparison of the SQ and ZQ line widths for the different crystal structures investigated.
    For the ZQ line, the uncertainty is shown based on the standard deviation over the 100 randomw structures.}
	\label{fig:SI_ZQSQgeneral}
\end{figure*}

We extended the line width and spin diffusion simulations from silicon to general crystal structures with the gyromagnetic ratio in the simulations set to $1$\,MHz and the lattice constant set to $1$\,\AA. 
Specifically, we simulated simple cubic, body-centered cubic (BCC), face-centered cubic (FCC) and diamond cubic crystal structures.
The cut-offs for the dipolar interaction as shown in Fig.~\ref{fig:SI_dipolarCutoffGeneral} take a very similar form for all crystal structures.
Thus, it is not suprising that the nuclear line single- and zero-quantum (SQ and ZQ) line widths take a similar appearance as those encountered in the main text (cf. Figs.~4 of the main manuscript and \ref{fig:SI_ZQSQgeneral}).

\clearpage

\section{Relaxation time variation in build-up and decay simulations}

\begin{figure*}[ht]
	\centering
	\includegraphics[width=\linewidth]{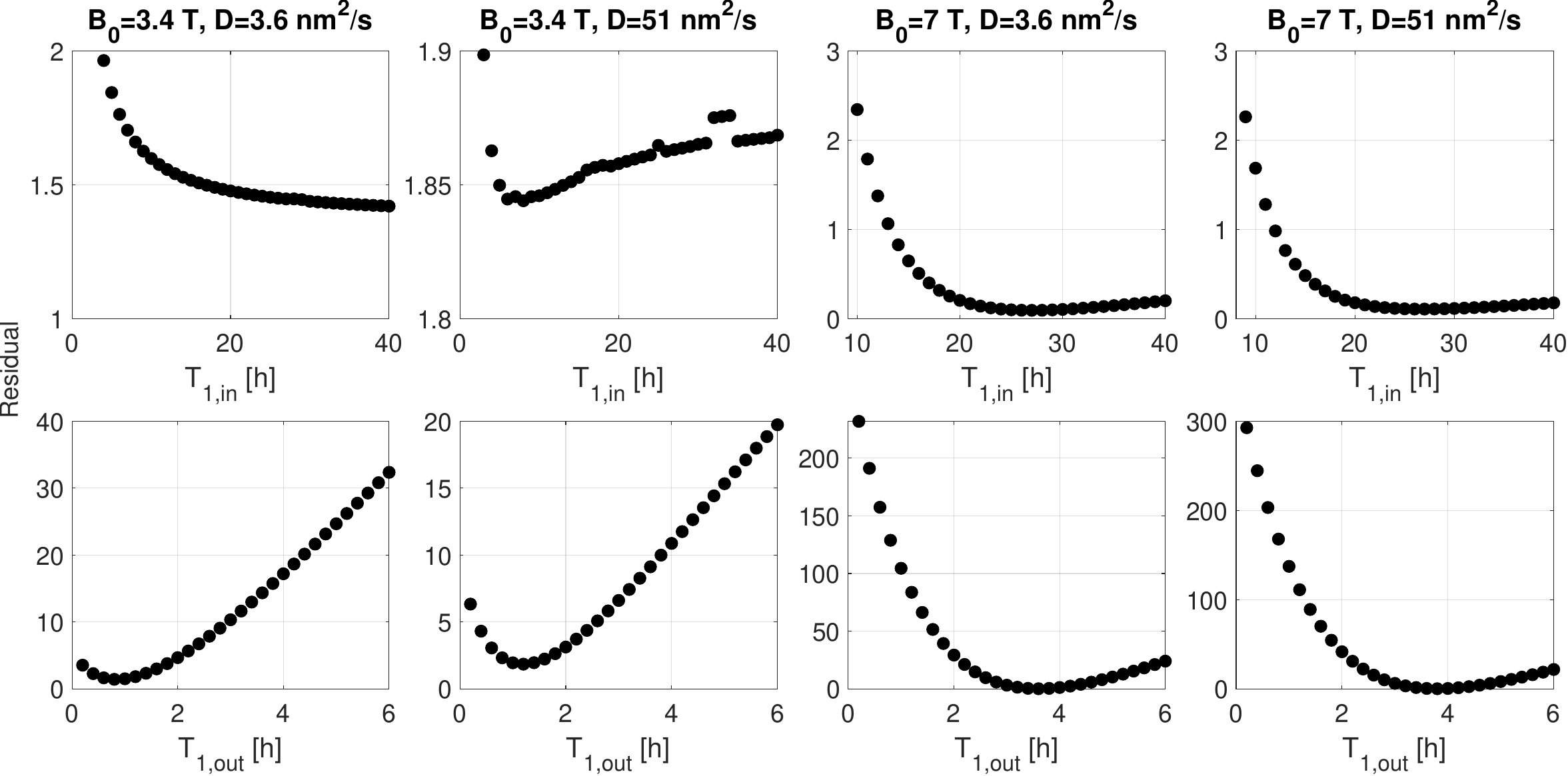}
	\caption{Least squares residue for variation of the relaxation times $T_\mathrm{1,in}$ and $T_\mathrm{1,out}$ of the build-up simulations.
    One of the two relaxation times is kept at its best fit value while the other is varied. 
    For the spin diffusion coefficient $D$, the 4.7\% (natural) \textsuperscript{29}Si abundance value was used, calculated with the nearest neighbor (Eqs.~4 of the main manuscript, $D=3.6$\,nm\textsuperscript{2}/s) or lattice (Eq.~6 of the main manuscript, $D=51$\,nm\textsuperscript{2}/s) model.
    The simulated build-ups have a higher sensitivity to $T_\mathrm{1,out}$ than $T_\mathrm{1,in}$ (cf. the change of residue).
    The better quality of the 7\,T measurements (more data points and higher SNR) translates into better fits (cf. main part, especially Fig.~6a of the main manuscript).
    The independence of the 7\,T fits from the spin diffusion coefficient and the better fit quality suggests that the observed build-up times are independent of the spin diffusion as for both values of $D$ the diffusion time is much shorter than the build-up time (seconds vs. hours).}
	\label{fig:SI_bupT1variation}
\end{figure*}




\begin{figure*}[ht]
	\centering
	\includegraphics[width=\linewidth]{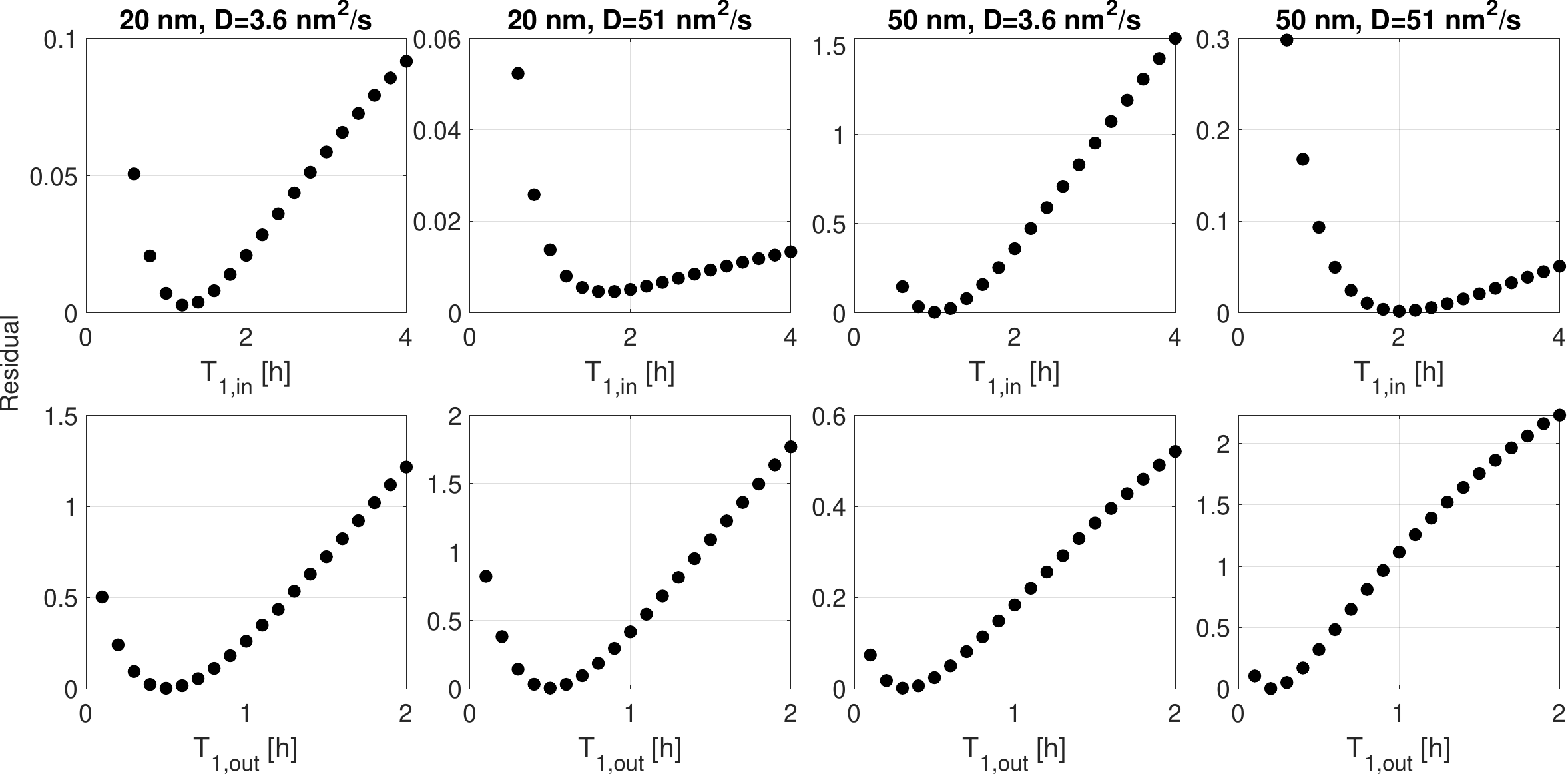}
	\caption{Least squares residue for variation of the relaxation times $T_\mathrm{1,in}$ and $T_\mathrm{1,out}$ of the decay simulations.
    One of the two relaxation times is kept at its best fit value while the other is varied. 
    For the spin diffusion coefficient $D$, the 4.7\% (natural) \textsuperscript{29}Si abundance value was used, calculated with the nearest neighbor (Eqs.~4 of the main manuscript, $D=3.6$\,nm\textsuperscript{2}/s) or lattice (Eq.~6 of the main manuscript, $D=51$\,nm\textsuperscript{2}/s) model.
    The simulated decays have a higher sensitivity to $T_\mathrm{1,out}$ than to $T_\mathrm{1,in}$ (cf. the change of residue).
    The fits for the 20 and 50\,nm particles appear similar.
    For both particle sizes, the different spin diffusion coefficients lead to similar results with the relaxation time on the outside shorter than on the inside.}
	\label{fig:SI_decayT1variation}
\end{figure*}

\clearpage

\bibliography{references,notes}